\documentstyle[preprint,tighten,pre,aps, epsfig]{revtex}

\begin{document}

\draft

\title{Velocity statistics in excited granular media}

\author{W. Losert$^{1}$, D.G.W. Cooper$^{1}$,
J. Delour$^{1}$, A.Kudrolli$^{1,2}$, and J.P. Gollub$^{1,3}$ \\}

\address{$^{1}$Department of Physics, Haverford College, Haverford PA 19041,
U.S.A.\\ $^{2}$Department of Physics, Clark University, Worcester MA 01610,
U.S.A.\\ $^{3}$Physics Department, University of Pennsylvania,
Philadelphia PA 19104, U.S.A}

\author{To appear in {\it Chaos}, September 1999}

\maketitle

\begin{abstract} We present an experimental study of velocity
statistics  for a 
partial layer of inelastic colliding beads driven by a vertically 
oscillating boundary.
Over a wide range of parameters
(accelerations 3-8 times the gravitational acceleration),
the probability distribution $P(v)$ deviates measurably 
from a Gaussian for the two horizontal velocity components.
It can be described by
$P(v) \sim \exp(-|v/v_c|^{1.5})$,
in agreement with a recent theory. The characteristic velocity $v_c$
 is proportional to the peak velocity of the boundary. 
The granular temperature, defined as the mean square particle velocity, varies with 
particle density and exhibits a maximum at intermediate densities.
On the other hand, for free cooling
in the absence of excitation, we find an exponential velocity distribution.
 Finally, we examine the  sharing of energy 
between particles of different mass.  
The more massive particles are found to
have greater kinetic energy.

\end{abstract}


 We determine the 
statistical properties of particles in a vibrated granular medium 
experimentally. While many similarities between ordinary
gases and excited granular media have been found, 
a fundamental difference is that collisions between particles
in granular matter are inelastic. As a consequence, 
the velocity distribution deviates measurably from a Gaussian, but 
can be described by $P(v) \sim \exp(-|v/v_c|^{1.5})$ for a large range
of parameters where external excitation is sufficiently frequent,
in agreement with a recent theory. 

\section{Introduction}

Granular systems are often treated statistically, since the large number of
degrees of freedom and the complexity of interparticle  forces 
limit analyses on the particle level.
What are the statistical properties of particles in an excited 
granular medium? In many
respects the dynamics of granular media strongly resemble the 
dynamics of ordinary fluids and solids. 
Yet there are fundamental 
differences between granular solids and fluids and their molecular 
counterparts that make the answer to this question 
intriguing~\cite{campbell90,jaeger96}.
Ordinary temperature does not lead to measurable velocity fluctuations 
due to the large mass of granular particles. When granular particles
 are excited by an external energy source,
their thermal energies are much smaller than their kinetic energies.
During collisions between particles, 
frictional forces and deformation near the points of contact lead to
dissipation of kinetic energy.
 Therefore it is not surprising to find that kinetic energy is 
transformed into thermal energy when particles interact in 
an excited granular medium~\cite{herrmann93}. 

For strongly excited granular media, assumptions similar to
those in kinetic theory are often made~\cite{campbell90}.  
Among these assumptions
are that velocity distributions are Gaussian and that the mean energy
is shared equally among the various degrees of freedom. However,
recent numerical and theoretical research indicates that 
excited inelastic hard spheres can exhibit non-Gaussian 
velocity distributions 
\cite{Goldhirsch96,Poeschel97,Noije98,Vulpiani98a,Vulpiani98b,Ohta98,Ruiz99},
though the predicted velocity distributions differ considerably
from each other. Some thermodynamic descriptions of granular media
make use of the concept of entropy~\cite{edwards91}, or separate the
dissipative degrees of freedom from conservative ones~\cite{herrmann93}.

Vibrated granular media have played a special role in efforts to
understand the dynamics of granular materials,
 in part because vibration is a convenient method
of replacing the energy lost to friction and inelasticity. 
A variety of novel phenomena have been discovered 
in the past fifteen years
\cite{evesque92}: heaping and convection rolls
\cite{evesque89,ehrichs95}, standing and traveling waves
\cite{douady89,pak93,melo95}, oscillons~\cite{Umbanhowar96},
and fluidization~\cite{clement93,luding94,clement91}.

In this paper, we study the simple case of a single layer of
particles with vertical excitation, varying the fractional coverage
of this layer and the amplitude of the driving waveform.  We
determine the velocity distributions precisely and compare them to various
functional forms, an issue that previous experiments have left unresolved for
granular particles that are free to move in three dimensions. 
Quantitatively studying particles in three dimensions 
has only recently been achieved in rotated cylinders 
using NMR techniques; it yielded important information about the 
segregation process~\cite{Nakagawa93}. By observing a three dimensional
excited granular medium from above, we are able to focus on the shapes of the
distributions of the horizontal velocity components.
A combination of otherwise identical
white and black particles allows us to track individual particles
even for high fractional coverages at high excitations where
particles frequently move over each other.
We measure the dependence of the variances of the distributions 
on the excitation amplitude, frequency 
and the fractional coverage. Sharing of energy
in mixtures of granular materials  is also investigated.
  
In Sec.~II, we discuss the background for our experiments.
In Sec.~III we present the experimental setup. 
In Sec.~IV,
we show that, in most cases we treated experimentally, the
velocity distribution $P(v)$ deviates measurably from a Gaussian and can 
be described by $P(v) \sim \exp(-|v/v_c|^{1.5})$.
We show precisely how the variance increases with
driving amplitude and changes with fractional coverage 
$c$, where $c=1$ is the coverage for a compact crystal of beads.   
The energy sharing between particles of different types 
is also described here.

\section{Background}

A number of studies on excited granular media have addressed
the issues of clustering and energy sharing. 
Numerical simulations and theoretical derivations have
shown that the presence of inelasticity in granular flows can lead to
the formation of clusters; as a consequence, equipartition of
energy fails \cite{louge90,goldhirsch93,du95}. Experiments performed
in a horizontal two-dimensional layer were consistent with this
predicted clustering effect \cite{arshad97}. Vertical one-dimensional
experiments and simulations were performed earlier \cite{luding94}; a
crossover from a condensed (clustered) to a fluidized state was found as a
function of the driving acceleration, the number of beads, and the
coefficient of restitution. 

Deviations from equipartition due to
clustering are straightforward to understand on physical grounds.
Inelastic collisions imply a loss of energy each time a collision occurs.
When particles begin to gather in a certain region of space, the rate of
their collisions increases. The rate of energy loss for this group of
clustered particles is thus greater, and the distribution of their
velocities becomes narrower than that of the particles in less 
dense regions. 

Clustering can affect the velocity statistics. 
Assuming a Gaussian velocity distribution for a nearly
homogeneous granular medium, Puglisi 
{\it et al.}~\cite{Vulpiani98a,Vulpiani98b} 
predict a non-Gaussian velocity distribution due to clustering:
a superposition of Gaussian velocity distributions with 
different widths. For inelastic particles this leads to 
high velocity tails in the velocity distribution
which decrease more slowly than a Gaussian function
but faster than an exponential.
The velocity distributions obtained in a simulation by Taguchi and
Takayasu~\cite{Takayasu95} are power laws, resulting
from clustering. These distributions have diverging variance; 
this calls into question the notion of a granular temperature, which is 
generally derived from a variance.
In a two-dimensional simulation, Peng and Ohta~\cite{Ohta98}
 found that the velocity 
distributions deviate from Gaussian behavior under the influence
of gravity unless $g \delta h << T$, where $\delta h$ is the height 
of the region of observation. 
$T$ is the granular temperature (see e.g. Ref.~\cite{campbell90}), 
defined as the variance 
of the velocity (minus the mean velocity)
\begin{equation}
T=<(v^2-<v>^2)>~.
\end{equation}

The high energy tails of velocity distributions in a homogeneous
granular fluid were investigated theoretically 
by Esipov and P\"oschel~\cite{Poeschel97} for the unforced case, and by
Noije and Ernst~\cite{Noije98}
for both the unforced and the heated case, 
based on the Enskog-Boltzmann equation.
In the free case, i.e. without energy input into the system, 
the velocity distribution approaches an exponential at high velocities.
When energy is added to the granular medium randomly and uniformly throughout
the system, the high velocity approximation
becomes $P(v) \sim \exp(-|v|^{1.5})$.
In order to compare our experimental results to this prediction 
we have to assume that energy input into the measured horizontal motion
occurs randomly.

While measurements of velocity statistics in an excited 
granular medium have been carried out~\cite{warr95}, few
 measurements precise
enough to distinguish between different functional forms
 of the velocity distribution
exist to our knowledge. One exception
is a recent study of clustering and ordering near a peak acceleration 
of $a=1$~g by Olafsen and Urbach~\cite{Urbach98}, who found 
significant deviations from Gaussian distributions, at low and 
especially at high velocities, where the distributions become exponential. 

Deviations from equipartition can occur for reasons other than clustering.
For example, Knight and Woodcock \cite{knight96} studied a
vibrationally excited granular system theoretically (without gravity)
and concluded that equipartition need not be observed at high amplitude
of excitation due to the anisotropy of the energy source. A two
dimensional system (a vertical Hele Shaw cell) was studied
experimentally by Warr, Huntley, and Jacques \cite{warr95}.  Velocity
distributions, though roughly Gaussian,  were checked and found
to exhibit anisotropy between the vertical and horizontal motion: the
horizontal velocity distribution was narrower than the vertical one.
Grossman, Zhou, and Ben-Naim \cite{grossman97} considered a
two-dimensional granular gas (without gravity) 
from a theoretical quasi-continuum point
of view and found the density to be nonuniform and the velocity
distributions to be asymmetric for ``thermal" energy input from one
side.  McNamara and Luding \cite{mcnamara98} considered the sharing of
energy between rotational and translational motion of the particles,
and found a significant violation of equipartition.

The scaling of the granular kinetic energy with vibration
amplitude is also of interest in connection with the experiments to be
discussed in the present paper.  It has been considered experimentally
by Warr et al.\cite{warr95}, numerically by Luding,
Hermann, and Blumen \cite{luding94b}, and theoretically by Kumaran in
the nearly elastic limit of weak dissipation \cite{kumaran98} and also
 by Huntley \cite{huntley98} in a simple model.  The
results of these different studies do not seem to be mutually consistent
with each other, perhaps because different regimes were explored; the
situation is unclear.

\section{Experimental setup and methods}

The experiments are conducted in a circular container of diameter 
32~cm made of delrin. It is driven vertically with sinusoidal
acceleration at a single frequency \cite{arshad96}
using a VTS500 vibrator from Vibration Test Systems Inc.
A computer controlled feedback loop keeps 
the vibration amplitude constant and reproducible.
The frequency $f$ used in most experiments is $100$~Hz 
and the peak acceleration $a$ of the plate 
is in the range $3-8~g$. The peak plate velocity $v_p=a/2 \pi f$ lies 
between $4.7$~cm/s and $14.1$~cm/s.
 The particles are
glass beads $4$ mm in diameter (from Jaygo Inc.) 
with fractional coverage $c$.  
A glass cover at $2$ cm height
prevents the beads from escaping from the
container. Collisions with the cover are rare for most experimental
parameters but lead to 
measureable changes in the velocity distributions at the 
largest accelerations, if the coverage is low. Some charging of
the glass beads is noticeable when the beads are at rest, but 
electrostatic forces are negligible at the range of accelerations 
investigated here.

Our objective is to measure the horizontal velocity distributions 
in an excited three dimensional granular medium for a large range 
of particle densities. 
However, particle tracking becomes increasingly difficult as particle
tracks approach each other and cross frequently  at large $c$, 
thus hindering reliable identification
of the horizontal positions of individual particles.
However, if only a modest number of particles are reflective,
the frequency of collisions between them
is small and nearly independent of coverage or excitation.
By tracking these test particles
 we can also directly compare the velocity distributions for
different coverage.

Tracking a subset of the particles
 is accomplished by using some white glass beads 
 among black glass beads. Except for the color, the black and white 
beads have identical physical properties. 
Stainless steel beads of three different diameters replaced the 
white glass beads in some experiments. The material properties
of all particles we used are listed in table~\ref{mat-table}.

Images of an area $16.74 \times 15.70$~cm at the center of the
container are taken at a resolution of  $512 \times 480$ pixels using a
 fast camera (SR-500, Kodak Inc.) operated at $250$ or $500$~frames/s.
At a vibration frequency of 100 Hz this ensures that images are taken
at 5 different phases relative to the phase of the plate vibration
yielding the average energy throughout the cycle. 
The images are analyzed using IDL (Research Systems Inc.) software. 
First each image is enhanced using a bandpass filter and 
thresholding to eliminate noise. The positions of all bright particles
are found from the enhanced image by calculation of
their centroid; this defines the particle positions reproducibly to within
less than $0.1$~pixels. Small effects due to the finite pixel
size are noticeable
sometimes as slightly increased probabilities of 
particle displacements that
are multiples of the pixel width ($0.0327$~cm in the physical system).
This displacement corresponds to $v=8.18$~cm/s in most experiments. 
The high frame rate ensures that even the fastest
beads move less than one particle diameter between images. This
allows accurate tracking of all bright beads for all $546$ sequential frames 
(the maximum available with our camera) with a typical precision
of $\pm 2\%$ for the velocity measurements. We project
each step onto two perpendicular directions,
and study the statistics of each velocity component.
One concern was that as a black bead moves over a bright bead,
the centroid position could move away from the center of the bead
and possibly alter the measured velocity distribution.
For a test we thus eliminated particles whose integrated greyscale intensity 
changed rapidly. We found that eliminating these points does 
not measurably alter the velocity distribution.
However, if a collision occurs between frames, our measurements indicate
the average of the velocity prior to and after a collision.
The measured distribution of velocities will therefore probably 
be slightly closer to a 
Gaussian than the real distribution of velocities.  
To improve the data in this regard, it would be necessary to
sample faster while retaining the same relative accuracy in the 
velocity measurement. It would therefore be necessary to use a faster camera,
to zoom in closer to the sample and to extend the measurement 
over a significantly larger number of frames to obtain the same statistics.

\section{Experimental Results}

\subsection{Granular Temperature}

\begin{figure}
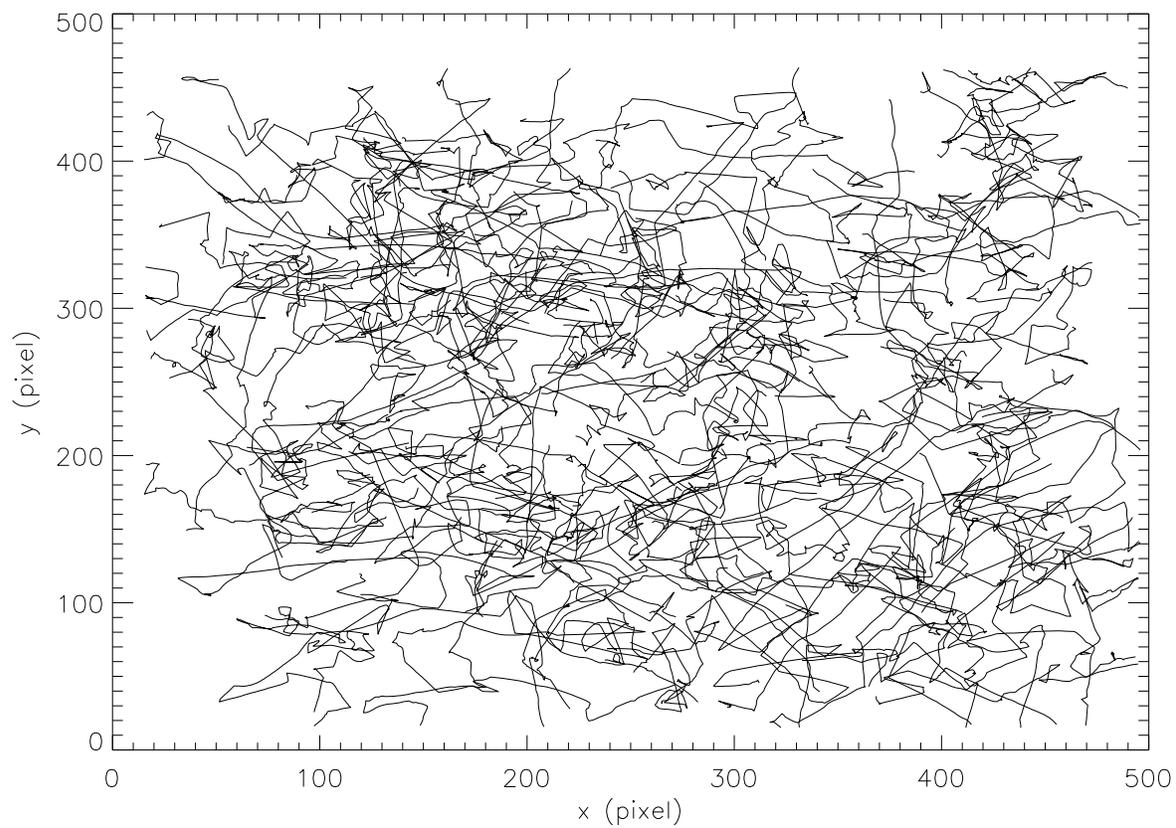
 
\caption{Measured trajectories of some  white particles 
moving among a sea of dark ones at a total coverage $c=0.28$. 
The container is vertically vibrated at 100 Hz and peak acceleration 
5 g. 
We acquire $546$~frames at $250$ frames/s. 
The particle diameter is about $11$~pixels).} 
\label{tracksfig} \end{figure}

Extracted particle tracks are shown in Fig.~\ref{tracksfig}
for $c=0.28$ (half of the beads are white) 
and $a=5~$g. 
For clarity, only tracks longer than $200$ images are shown, 
which eliminates some tracks close to the edge.  
A total of $546$ frames is acquired at $250$~frames/s 
(i.e. for $2.18$~seconds) with approximately $200$ 
tracked particles in each frame.

\begin{figure}
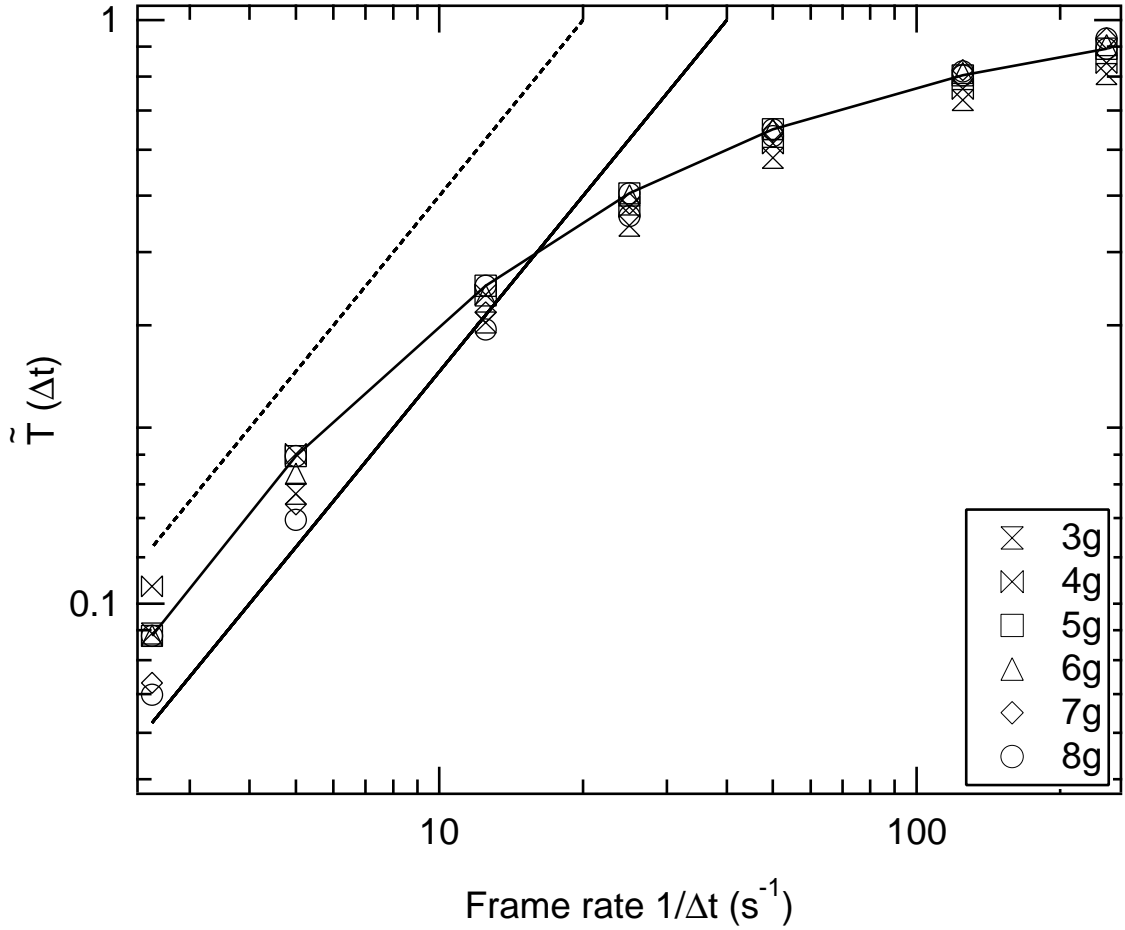
 
\caption{Apparent non-dimensional granular temperature $\tilde{T}( \Delta t)$
corresponding to displacements along one axis
in the sample interval as a function of sampling rate $1/ \Delta t$.    
The approach to saturation for large sampling rates occurs when the
sampling rate far exceeds the collision rate. In the opposite limit
$\tilde{T}$ decreases strongly. The lines are explained in the text. }
\label{tempvsframesfig} \end{figure}

The particle velocities are determined from  the particle
displacement between consecutive frames. 
This does not always represent the true 
velocity of the particle though. If a collision occurs between frames, the 
apparent velocity will be lower than the true velocity. 
We can define an apparent temperature based on displacements 
along either horizontal coordinate denoted here as $x$; 
it depends on the time between frames $\Delta t$:
\begin{equation}
T(\Delta t)=<(x_j(t_k+\Delta t)-x_j(t_k))^2>_{j,k}/ \Delta t^2~,
\end{equation}
where the average is taken
over all particles (j) and frames (k). 
The particle tracks obtained at $250$~frames/s during an interval of $2.18$~s 
allow us to determine $T(\Delta t)$ for 
$1/250 {\rm ~s} \le \Delta t << 2.18 {\rm~s}$.
It is often useful to express the velocities in units of the peak plate
velocity $v_p$, which yields a dimensionless temperature 
$\tilde{T}(\Delta t)= T(\Delta t)/ v_p^2$.
Fig.~\ref{tempvsframesfig} shows the dimensionless temperature 
$\tilde{T}(\Delta t)$
vs. frame rate ($1/\Delta t$) at $c=0.42$ and $f=100~$Hz for different 
accelerations.
For small frame rates, i.e. large $\Delta t$, $\tilde{T}(\Delta t)$ increases
approximately linearly with the frame rate 
$\tilde{T}(\Delta t) \sim 1/\Delta t$. 
This indicates that the particle motion may be described by 
an ordinary diffusion law when many collisions occur in the sampling interval 
$\Delta t$. 
Assuming such a diffusion process, one expects that 
\begin{equation}
\tilde{T}(\Delta t) \approx \frac{\tilde{T} \tau_c}{\Delta t}~~~~~
({\rm for}~ \Delta t >> \tau_c)~.
\end{equation}
The dashed and the solid lines in Fig.~\ref{tempvsframesfig}  are linear
in $1/\Delta t$ and give  upper and
 lower limits to $\tilde{T} \tau_c$. 
We estimate that $\tilde{T} \approx 1.0$ based on
the high frame rate limit. We can now estimate the collision
time to be $ 0.02~{\rm s} < \tau_c < 0.04~{\rm s}$ for
 $f=100$~Hz and $c=0.42$,
which corresponds to one collision every $2-4$ oscillations of the vibrator.
On the other hand, for large frame rates (small $\Delta t$), 
$\tilde{T}(\Delta t)$ approaches a constant.
This occurs when $\Delta t$ is much smaller than $\tau_c$; 
 in this limit $T(\Delta t) \approx T$,
so the granular temperature $T$ (for displacements along 
one axis) can be defined as
\begin{equation}
T=\lim_{\Delta t\to0}T(\Delta t) = <v^2>~.
\end{equation}
The approach of the measured granular temperature 
to a constant cannot be fitted by a simple exponential or power law.
For an excited granular material, this shape could be influenced by
correlations in velocity between
neighboring particles and correlations between the local density 
and particle velocity. Starting from the fact that the sampling time 
is roughly $10\%$ of the mean collision time at an intermediate coverage,
we estimate that the true granular temperature 
$T$ might be up to $10\%$ higher than the measured value.

\begin{figure}
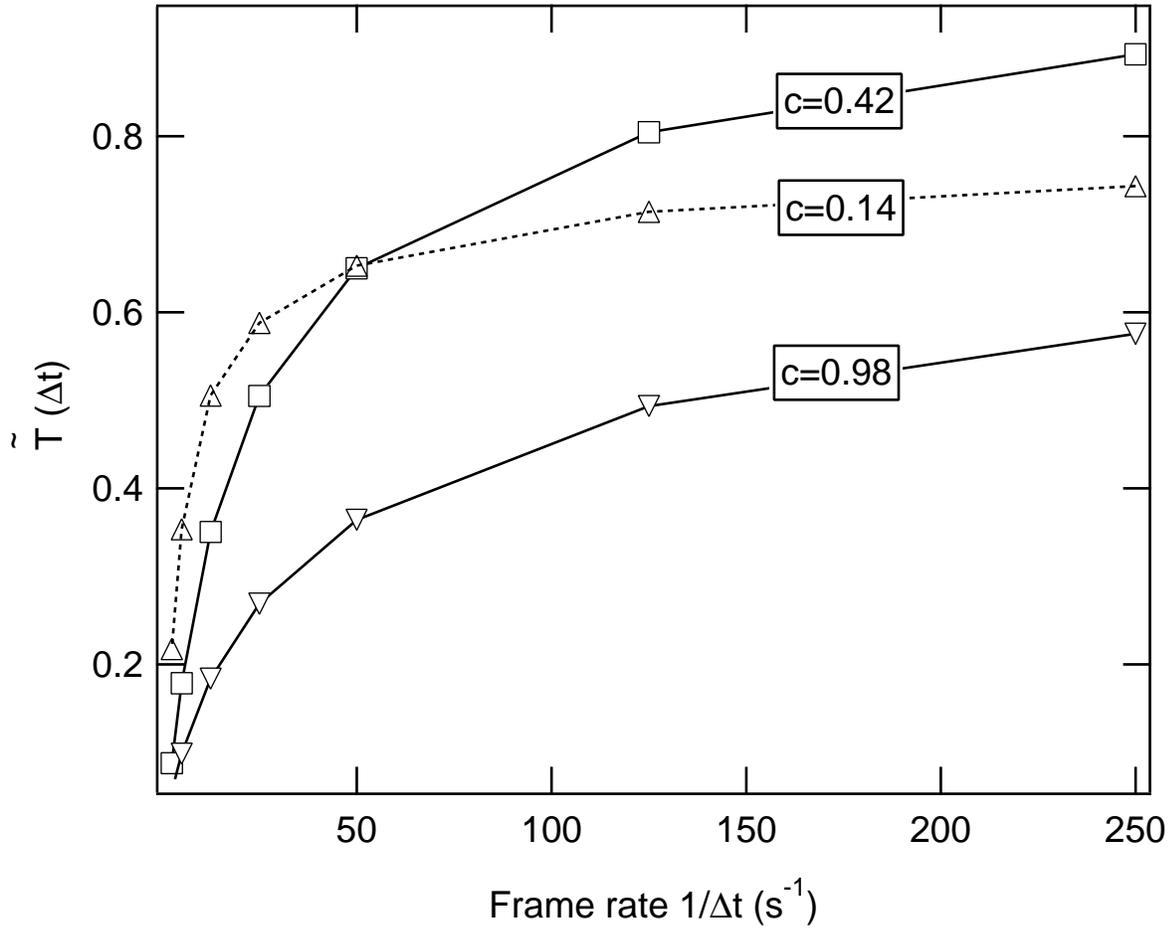
 
\caption{Apparent non-dimensional granular temperature 
 $\tilde{T}(\Delta t)$ corresponding to displacements along one axis
in the sample interval $ \Delta t$
as a function of sampling rate $1/\Delta t$ for $a=5$ g. 
The limiting value for large sampling rates depends on coverage.}
\label{fig3clabel} \end{figure}

All values of $T$ and $\tilde{T}$ presented in this paper are the measured
temperatures obtained at the {\it highest} frame rate 
(usually $250$~frames/s), where the limit
$\Delta t\to0$ is justified. This limit is approached in a very similar way 
for different accelerations (see Fig.~\ref{tempvsframesfig}).
This allows us to determine the acceleration dependence of $\tilde{T}$.
Note that the collision time can be roughly independent of acceleration
since the height of the average bounce increases with increasing
peak plate acceleration. For the fixed number of particles in our
system this leads to an increase in the mean free path with increasing
acceleration. When the number of particles is changed, the collision
time also changes. As shown in  Fig.~\ref{fig3clabel}, the limit of Eqn.~(4)
is approached fastest at the lowest energy of $c=0.14$ 
and slower at $c=0.42$ and at $c=0.98$. 
However, we are close enough to the limit at all coverages
(since the mean time between
collisions is always significantly longer than the time between frames) 
to observe qualitatively
how $\tilde{T}$ changes with $c$.

\begin{figure}
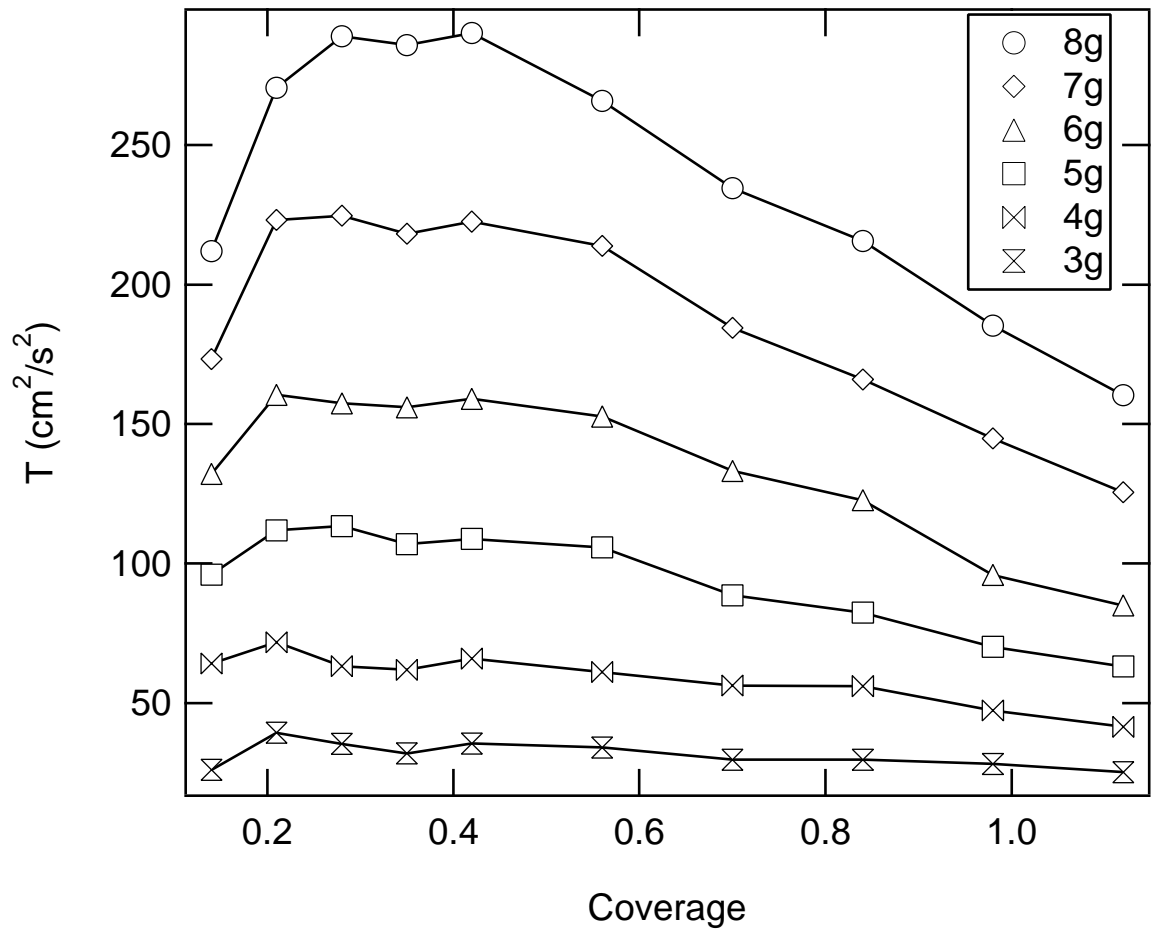
 
\caption{Granular temperature $T$ determined 
from displacements along one axis as a function
of coverage for a range of peak plate accelerations.} 
\label{tempvscovfig} \end{figure}

We can therefore determine the dependence of the
temperature on plate acceleration $a$ and coverage $c$, 
shown in Fig.~\ref{tempvscovfig}.
The temperature
increases with $a$ for all coverages. 
As a function of $c$, $T$ increases at low coverage,
exhibits a maximum around $c=0.30$ and decreases with $c$ at 
high coverage. 
This trend reflects changes in the true granular temperature
as shown in Fig.~\ref{fig3clabel}.
No measurable change in the temperature dependence
occurs around $c=1$. A fractional coverage above unity is meant to
indicate that more than one close packed layer of beads is used.
The highest granular temperature indicates that the average
potential energy of the particles corresponds to a mean height 
above the plate comparable to one particle diameter $d$. All experimental
results are therefore limited to the regime of particle energies smaller
than or comparable to the only characteristic energy of a granular material,
the potential energy of raising one particle by one diameter $m g d$.
When scaled by the peak plate velocity $v_{p}$ 
as in Fig.~\ref{tempvscovsclfig}, the granular temperature becomes 
independent of acceleration to within
$\pm10\%$ for most data points. Remarkably, the dependence on coverage 
follows approximately the same behavior at all accelerations.

The scaling of velocities by $v_{p}$  implies that $T \sim 1/f^2$.
Fig.~\ref{freqdepfig} shows, on a log-log plot, that $T$ does 
indeed decrease approximately $\sim 1/f^2$ for the two accelerations
and the three coverages shown. The plot covers two orders in magnitude 
of the granular temperature, ranging from conditions where
beads rarely hit the container lid to conditions where frequent collisions 
with the lid occur. We conclude that the scaling of the bead velocity by
the plate velocity is very robust and is not significantly 
influenced by additional 
contacts with the container lid.   

\subsection{Velocity Distributions}
 
The velocity distribution along one axis, obtained from particle tracks of 
 white beads, 
 is shown in Fig.~\ref{veldist-fitfig}.  
 Fits to a Gaussian distribution
$F_g(v)=A_g\exp(- |v/\sqrt{2} v_{c}|^2)$ 
are shown as dashed lines, 
and fits to the prediction of Ref.~\cite{Noije98}
\begin{equation}
F_2(v)=A_2\exp(- |v/ 1.164 v_{c}|^{1.5})
\label{equnonGauss}
\end{equation}
are shown as solid lines. 
The characteristic velocitiy $v_{c}$ is defined as the square root
of the variance, so that for both
$F_g$ and $F_2$:
\begin{equation}
v_{c} = \sqrt{<v^2>} = \sqrt{T}~.
\end{equation}
The data points for the fits are weighted equally on a linear scale  
in Fig.~\ref{veldist-fitfig}(a,b)  and equally on a logarithmic scale
in Fig.~\ref{veldist-fitfig}(c,d). 
The characteristic velocity $v_{c}$  in (a,b)
is $v_{c}=6.19$~cm/s for $F_g$ and  
$v_{c}=6.99$~cm/s for $F_2$.
The Gaussian fit underestimates the probability of both low and 
high ($v > 3v_{c}$) 
velocities, while the fit to $F_2$ describes the
probability distribution quite well over 
three orders of magnitude in probability. 
The increased weight of the high velocity experimental
data in (c,d) leads to  $v_{c}=7.66$~cm/s for $F_g$ and  
$v_{c}=6.81$~cm/s for $F_2$. The Gaussian fit again underestimates
high and low velocity probabilities, while the fit to $F_2$ proves
to be insensitive to the weighting of data points, indicating a robust
fit. The fit to $F_2$ is also insensitive to the
choice of the fitted range of velocities, while the characteristic velocity
decreases with decreasing fitting range for $F_g$.  We conclude that
$F_2$ provides a better fit than $F_g$.    

The velocity distribution for a large range of accelerations 
$3~{\rm g} \le a \le 8~{\rm g}$ can be described accurately
by $F_2$ as shown
in Fig.~\ref{veldist2fig}(a). 
The data and the best fitting lines are shifted vertically 
as needed for clarity; this amounts to multiplication by a constant on 
a log-linear plot. 
The probabilities are plotted against $|\tilde{v}|^{1.5}$, 
where $\tilde{v}=v/v_p$. 
On this log-linear scale, $F_2$ is a straight line, in 
good agreement with the experimental data for a large range of accelerations.
The dependence of the velocity distribution on coverage is more complex. 
The coverage dependence is shown in
Fig.~\ref{veldist2fig}(b). 
At relatively high coverage, above $c=0.28$, the experimental
data can be described well by $F_2$ (solid lines). 
However, at lower coverage (for approximately the same range of $c$
where $T$ increases with $c$) the probability
of high velocities is underestimated by $F_2$ (dashed line). 

While the granular temperature is proportional to $1/f^2$ to a good
approximation, measureable differences are apparent in the 
distribution of  non-dimensional velocities $\tilde{v}$, 
as a function of frequency. 
Fig.~\ref{veldist3fig} compares the distributions for $40$~Hz and $140$~Hz. 
The distribution falls more slowly with velocity as $f$ is increased.
The cause of this behavior is probably the smaller (unscaled) 
velocities at higher frequency,
which decreases the collision rate (i.e. the rate of energy loss
through inelastic collisions), and the higher frequency of vibration
(i.e. roughly the rate of energy input). For the
highest unscaled particles velocities and lowest excitation frequency 
($a=5$~g and $f=40$~Hz) deviations from fits to $F_2$ become observable.

At low accelerations $a=2$~g we observe that the velocity distribution
has exponential tails and an approximately Gaussian central component.
The crossover from a Gaussian distribution to an exponential distribution
is shown in Fig.~\ref{lowaccfig}. These results are similar to the velocity
distributions around $a=1$~g obtained by Olafsen and Urbach~\cite{Urbach98}, 
and possibly related to clustering effects observed at low accelerations. 
In addition, the system is nearly two dimensional since most beads bounce 
lower than one particle diameter..  

It is possible to investigate the case of a freely cooling
granular medium to some extent by looking at a time averaged 
velocity distribution. We shut off the vibrator abruptly, and simultaneously
trigger the camera at $250$~frames/s. 
We then extract particle tracks from the image sequence in the same
way as we did for image sequences of continuously
excited granular media.
In order to analyse the functional form of the velocity distribution, 
we need to accumulate velocities over $150$ frames, i.e. $0.6$~s.
Fig.~\ref{freecoolfig} shows such a velocity distribution for 
$c=0.84$, where the vibrator was operated
at $f=100$~Hz and $a=5$~g prior to shut off. 
The velocity distribution is exponential,
in agreement with calculations for the free cooling 
case~\cite{Poeschel97,Noije98}.
However, the measured  velocity distribution represents
an average over almost the entire free cooling process, since the
instantaneous velocities indicate that the 
granular temperature decreases by more than one order of magnitude
 during the averaging time. We have also made measurements over $0.1$~s;
although the statistics are not as good, the distributions still appear
to be exponential.  

\subsection{Equipartition}

In a final set of experiments, the white glass beads are replaced by
steel beads of different size 
(grade 100 stainless steel 316 for the two smaller sizes,
grade 100 stainless steel 302 for the largest beads).
At a total coverage $c=0.42$, $14\%$ 
of the glass beads were replaced by steel beads 
(for the smallest bead size only $5\%$ of the glass beads were replaced). 
In all experiments most collisions of steel beads
therefore occur with glass beads.
The distributions of $\tilde{v}$ for the tracked particles
are shown in Fig.~\ref{diffmass1fig}.
Larger beads have smaller non-dimensional characteristic velocities 
$\tilde{v_c}$ and thus a smaller granular 
temperature $T$ than smaller beads. 
The velocity distributions for the two smaller 
steel bead sizes are described well by $F_2$, while that
 for the largest beads is better described by a Gaussian 
(dashed line). This could indicate that the large beads effectively
prevent clustering since they act as a source of momentum for 
the other beads.
While the granular temperature of the largest steel beads is lower 
than the temperature of the glass beads,
their energy $m T$ is larger, while the energy of the smaller steel 
bead sizes is smaller, as shown in Fig.~\ref{diffmass2fig}. 

\section{Summary and conclusion}

We have reported experimental studies of
velocity statistics for a (fractional) layer of glass beads subjected
to  vertical vibration. The horizontal motion of a small subset 
of beads was measured using a high speed camera at a
frame rate sufficiently high to measure instantaneous velocities
accurately. The measurements were acquired over an interval substantially
longer than the time between interparticle collisions.
These capabilities allowed us to determine particle statistics 
of the horizontal motion in detail. 
We analyzed granular temperatures and velocity distributions 
for a large range of excitation frequencies, amplitudes, and 
coverages.

\begin{figure}
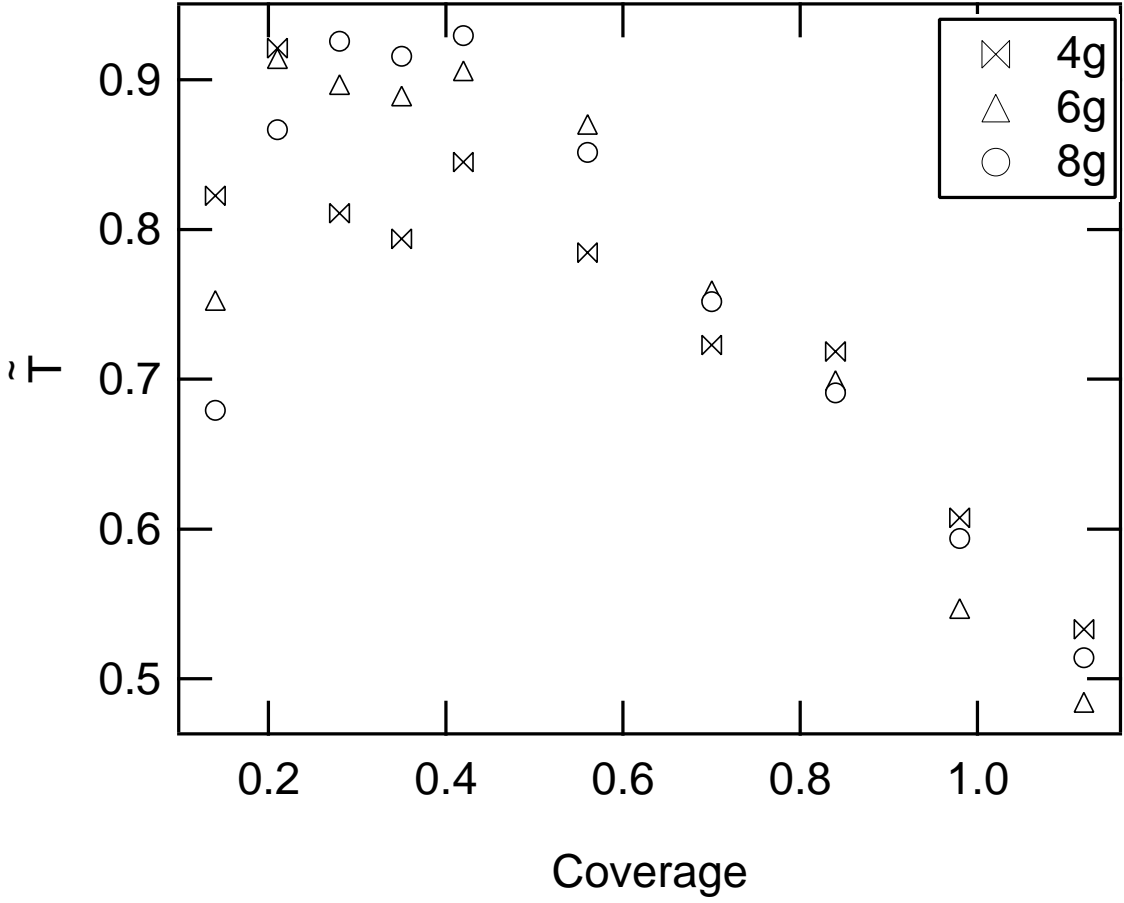
 
\caption{Non-dimensional granular temperature $\tilde{T}$ 
vs. coverage. It is approximately independent of acceleration.}
 \label{tempvscovsclfig} \end{figure}

\begin{figure}
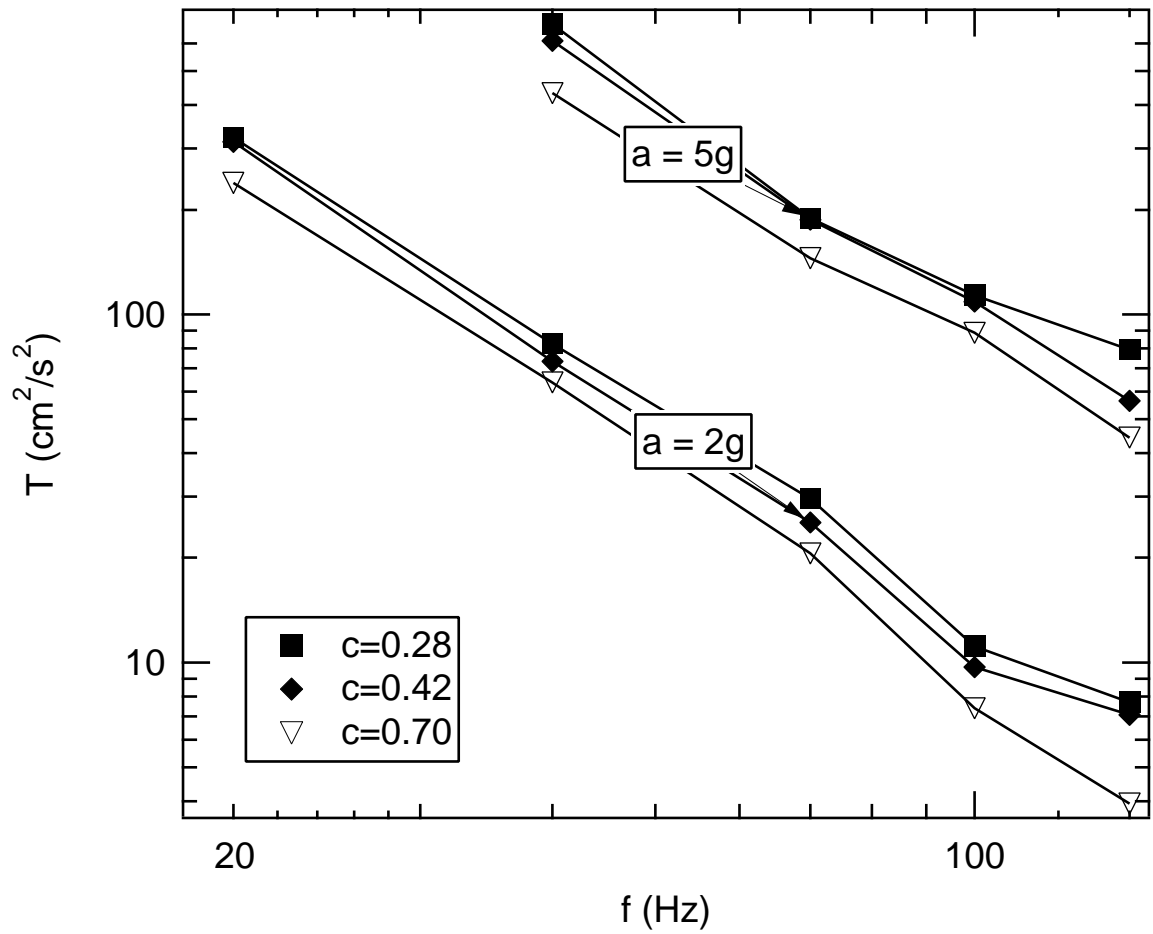
 
\caption{Granular temperature $T$ vs.
vibration frequency $f$ at three coverages. It declines
approximately as $1/f^2$.}
 \label{freqdepfig}  \end{figure}

The variance of the particle velocity distribution
(or the granular temperature $T$ of horizontal motion)
varies approximately in proportion to 
the plate velocity (Fig.~\ref{tempvscovsclfig}). 
It increases with increasing coverage $c$
at low $c$ and decreases at higher $c>0.42$. On the other hand,
the mean energy associated with the vertical motion probably declines with
increasing $c$ for all $c$ due to additional dissipation from 
collisions. 
In this interpretation the decrease in $T$ with $c$ at high
$c$ mirrors the decrease in energy associated with vertical motion.
In contrast, the smaller values of $T$ found at low $c$  likely indicate
that the energy transfer from vertical to horizontal motion becomes 
less efficient at low $c \le 0.42$.

We have shown that particles of different mass do not have the
same kinetic energy or the same granular temperature 
when both are present simultaneously 
(Fig.~\ref{diffmass2fig}), an
apparent violation of equipartition.  
The most reasonable explanation is that all particles acquire similar
vertical velocity fluctuations from the container.  The more massive
particles therefore obtain a larger vertical kinetic energy.
This excess vertical energy is transferred to the horizontal motion, leading
to a violation of equipartition. 

\begin{figure}
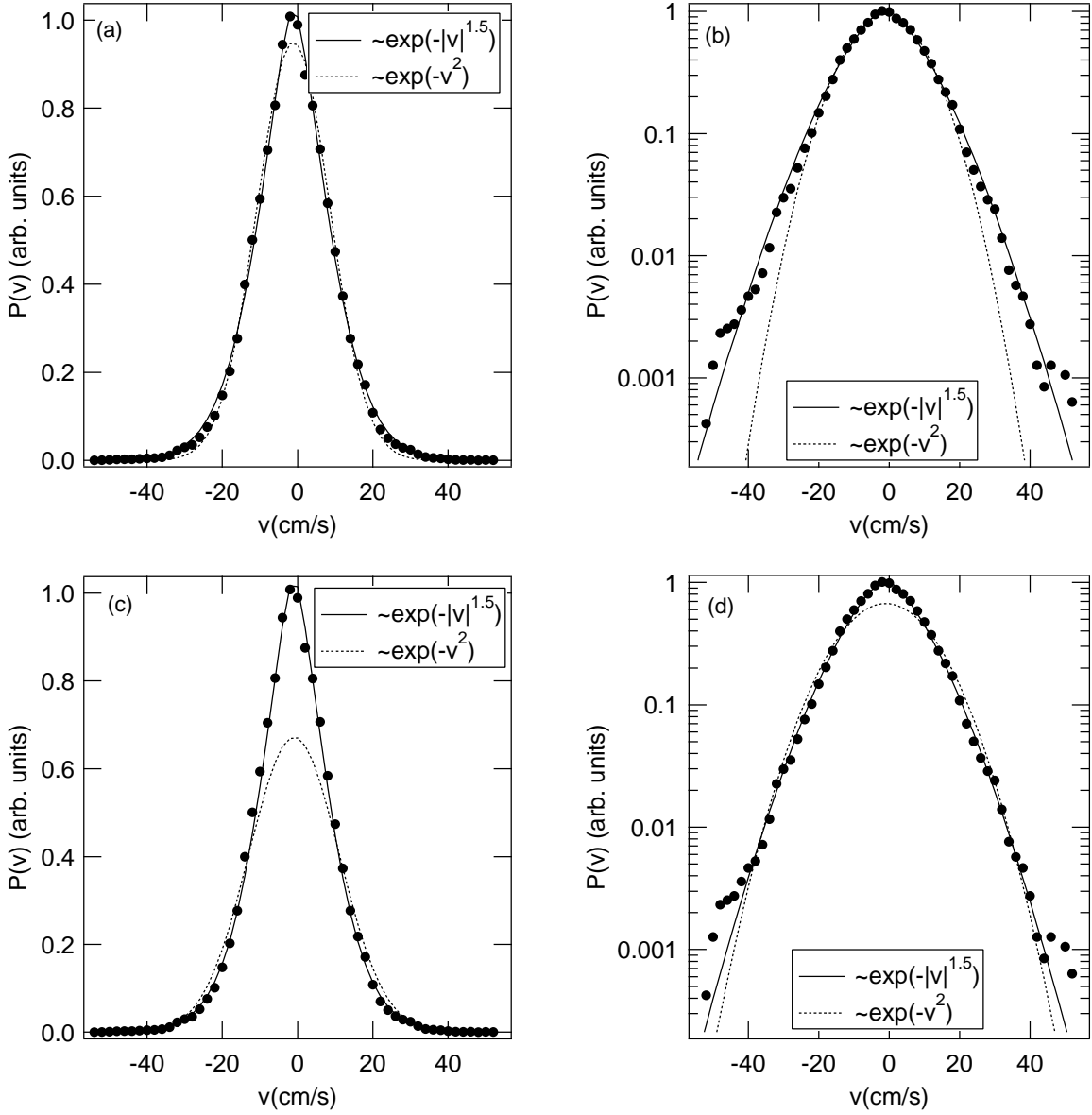
 
\caption{Velocity distribution of the tracked particles 
plotted on linear (a,c) and logarithmic (b,d) scales (driving acceleration 
$a=5$ g, coverage $c = 0.42$, and frequency $f = 100$~Hz). The solid line is a 
fit to $F_2$ (Eq.~\ref{equnonGauss}).
The Gaussian fit 
(dashed line) 
underestimates the 
probablilities at high velocities. The data were weighted 
equally on linear scales in (a,b) and equally on logarithmic scales 
in (c,d), respectively.} 
\label{veldist-fitfig} \end{figure}

\begin{figure}
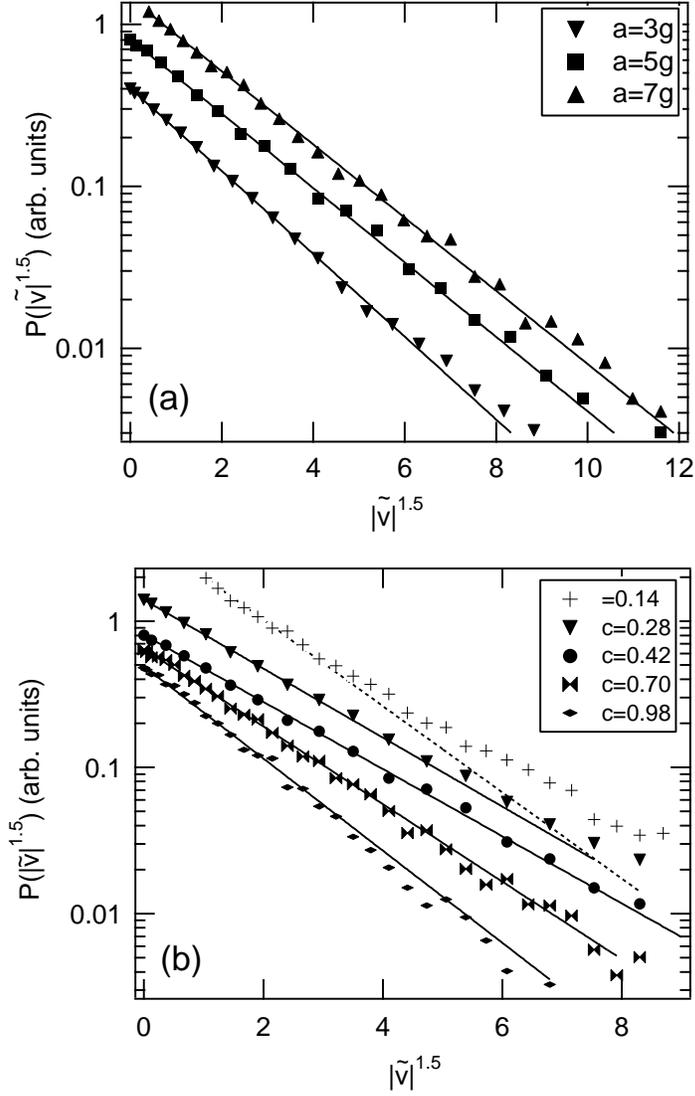
 
\caption{Distributions of non-dimensional 
velocity $\tilde{v}$ obtained from
displacements along one direction
 vs. $\tilde{v}^{1.5}$ for (a) a range of accelerations and
(b) a range of coverages. Fits to Eq.~\ref{equnonGauss} (lines) are
also shown. Data are shifted vertically in some cases for clarity. Deviations 
from Eq.~\ref{equnonGauss} occur for $c\le 0.28$.}
\label{veldist2fig} \end{figure}

\begin{figure}
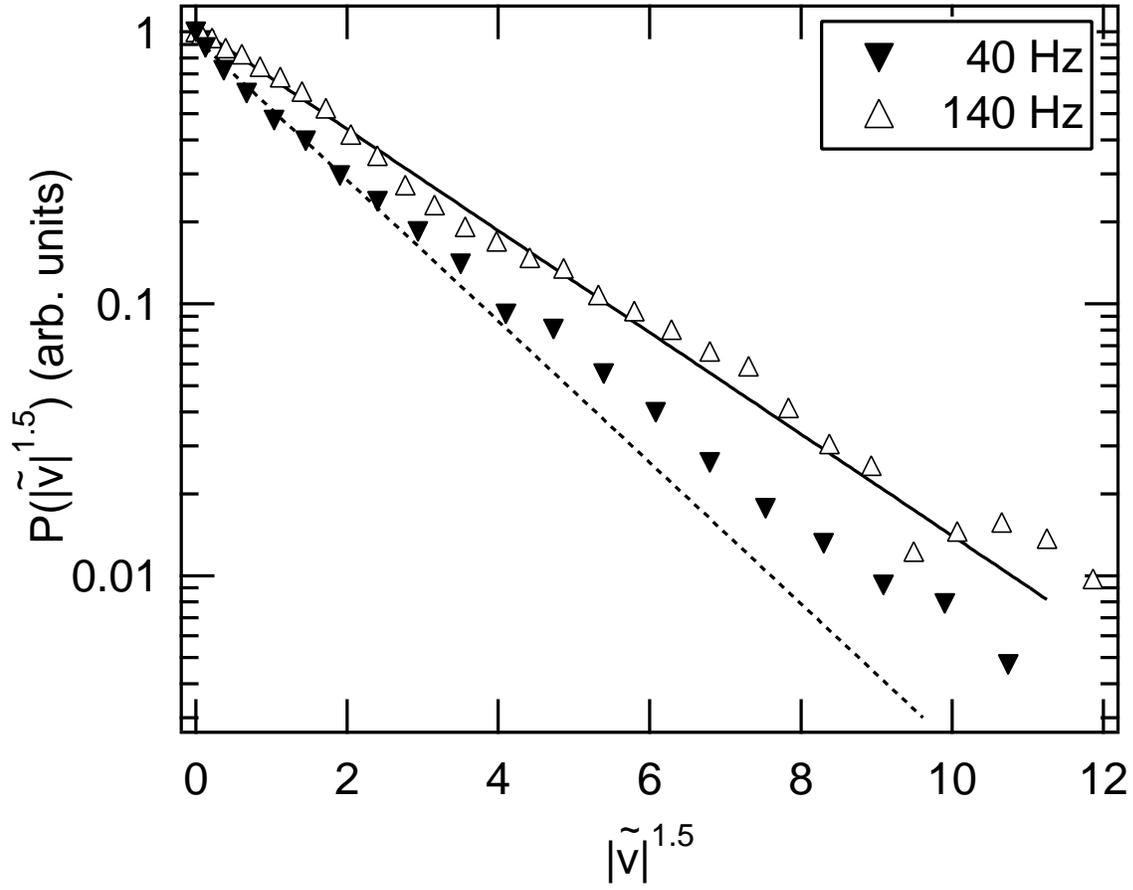

\caption{Distributions of non-dimensional 
velocity $\tilde{v}$ obtained from
displacements along one direction vs.
 $\tilde{v}^{1.5}$ at different vibration frequencies,
and fits to Eq.~\ref{equnonGauss}.
($c=0.42$, $a=5g$)}
 \label{veldist3fig}  \end{figure}

\begin{figure}
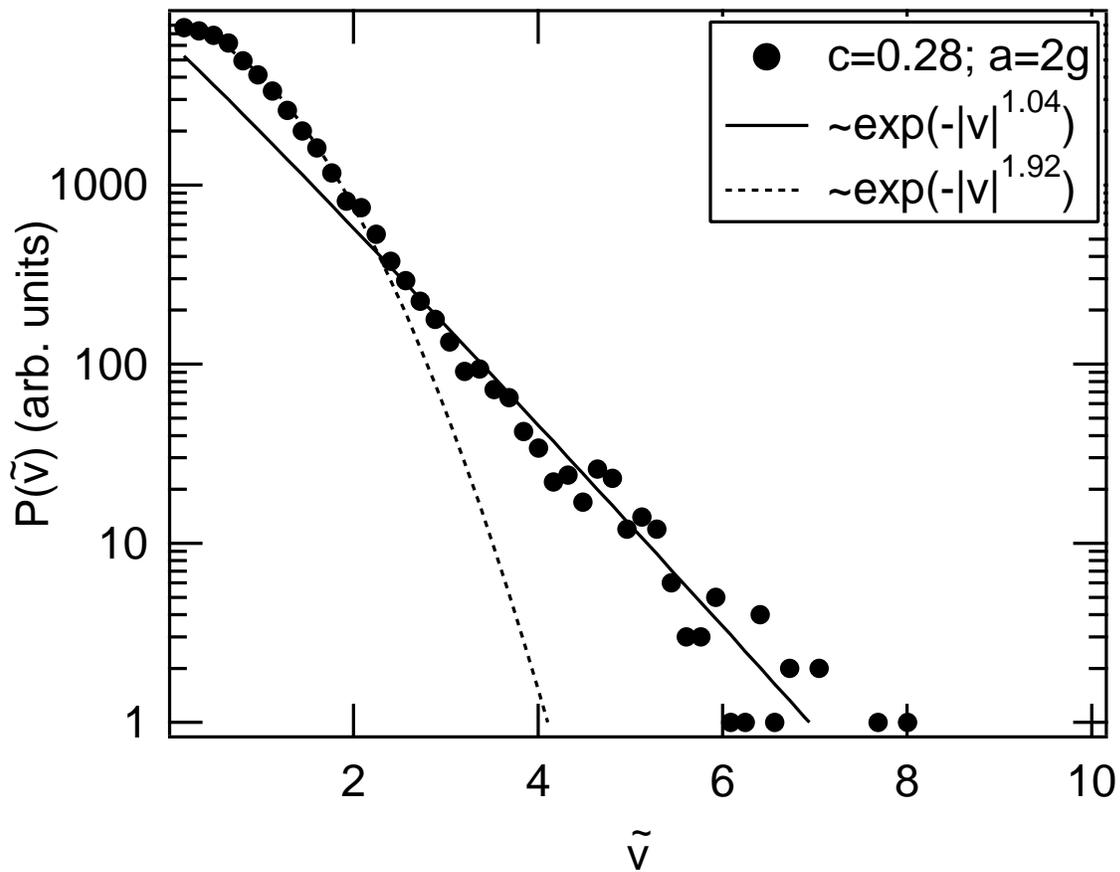
 
\caption{Non-dimensional velocity distribution at 
low acceleration $a=2$~g, and fits to
$P(\tilde{v}) \sim exp(-|v|^\alpha)$ 
showing the crossover from a Gaussian to an exponential function.}
 \label{lowaccfig} \end{figure}

\begin{figure}
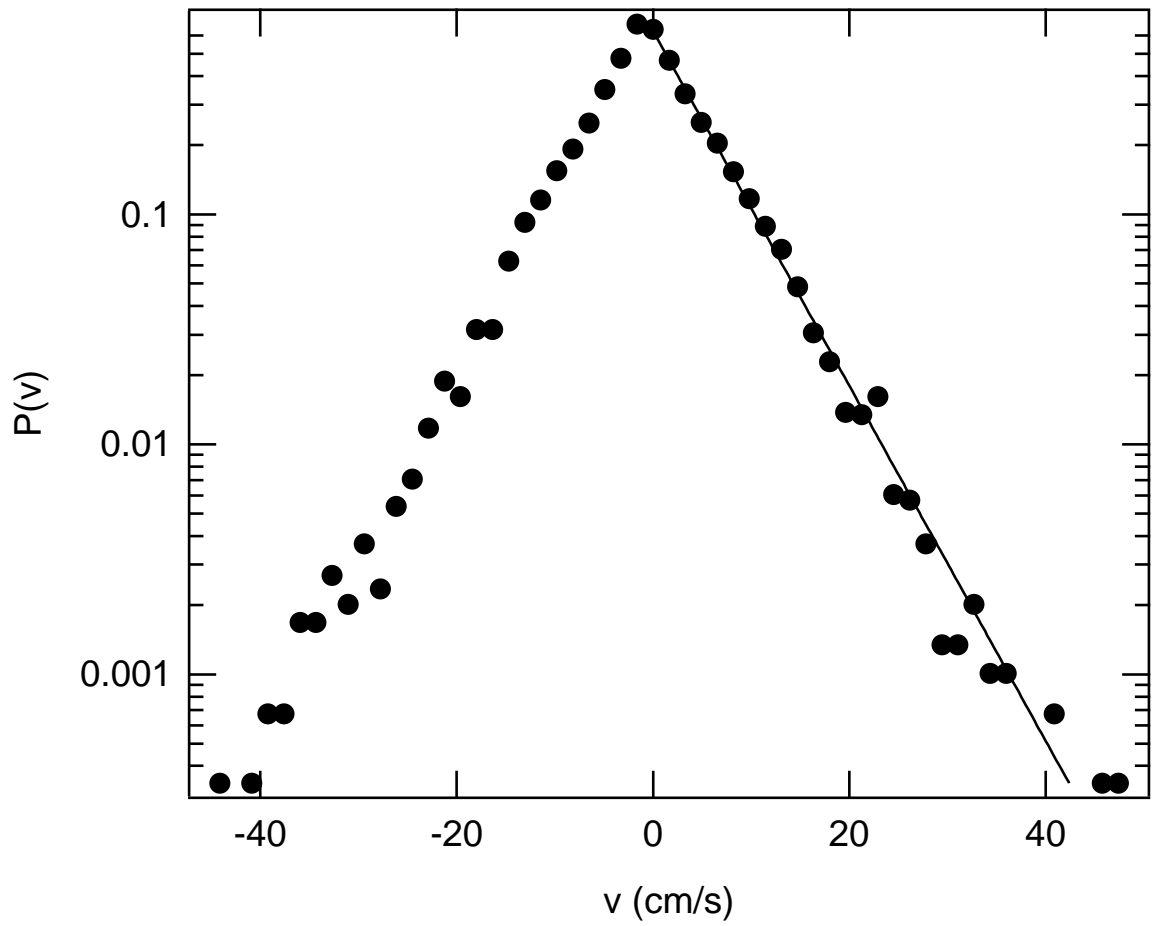
 
\caption{Time averaged velocity distribution
during free cooling after excitation at $a=5$~g and $f=100$~Hz ($c=0.84$).
The high velocity tail is exponential 
(a straight line on this log-linear plot).}
 \label{freecoolfig} \end{figure}

\begin{figure}
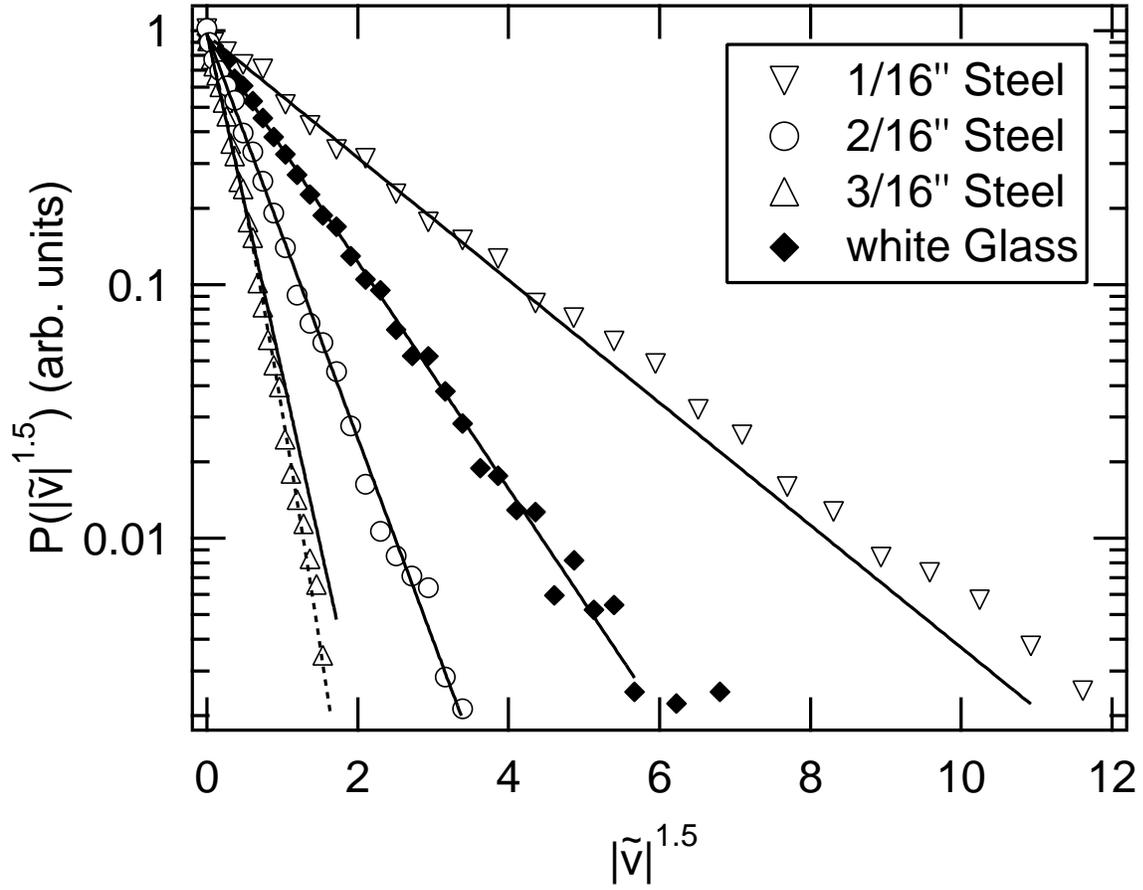
 
\caption{Velocity distributions for steel beads moving among 
glass beads ($c=0.42$, $a=4$~g, $f=100$~Hz). Distributions are
fitted to Eq.~\ref{equnonGauss}
(solid lines) and also to a Gaussian (dashed line) for the largest beads.}
 \label{diffmass1fig} \end{figure}

An important result of this investigation is that in
the steady state the velocity distributions deviate measurably from
a Gaussian (Fig.~\ref{veldist-fitfig}), but can be described well by 
$P(v) \sim \exp(-|v/v_c|^{1.5})$ for broad ranges of frequencies $f$,
accelerations $a$ and coverages $c$, in agreement with the theory of
Ref~\cite{Noije98}. In most of our experiments the time between 
interparticle collisions $\tau_c$ is comparable to the time between 
contacts with the plate.  
However, if forcing collisions become significantly less frequent than
 interparticle collisions (as for free cooling, Fig.~\ref{freecoolfig}), 
then the distribution approaches
an exponential. This experimental observation is consistent with the 
numerical results of Puglisi {\it et al.}~\cite{Vulpiani98a}. 
This quantitative correspondence with experiment indicates that the
theoretical and numerical approaches of Refs.~\cite{Noije98,Vulpiani98a}
to describing the statistical properties of granular particles are promising.

Puglisi {\it et al.}~\cite{Vulpiani98a} suggest that non-Gaussian behavior 
and clustering are indications of essentially the same particle dynamics.
If so, our experimental results indicate that clustering
must occur for a very large range of excitation amplitudes and frequencies.  

\textbf{Acknowledgments:}  This work was supported in part by the
National Science Foundation under Grant No. DMR-9704301.
We thank Eric Weeks and John Crocker for providing particle tracking software. 
We thank  Yuhai Tu for valuable discussions.
Technical support was provided by Bruce Boyes.



\begin{table} \begin{center}
\begin{tabular}{cccc}
Material & Diameter & Mass & Density \\
 &  [cm] & [g] & $[{\rm g}/{{\rm cm}^3}]$  \\
\hline
Glass (black and white) & 0.407 & 0.0864 & 2.46\\
Stainless Steel 316 & 0.159 & 0.0168 & 8.03 \\
Stainless Steel 316 & 0.318 & 0.1345 & 8.03 \\
Stainless Steel 302 & 0.476 & 0.4539 & 8.03 \\
\end{tabular}
\end{center}
\caption{Material parameters of granular particles. Black glass beads
are used as background particles in all experiments.}
\label{mat-table}
\end{table}

\begin{figure} 
\caption{Mean kinetic energy
of steel beads moving among glass beads (scaled by the energy of
the surrounding glass beads) as a function of particle mass.
($a=4$~g; $f=100$~Hz; squares, $c=0.42$; triangles $c=0.84$)}
 \label{diffmass2fig} \end{figure}

\setcounter{figure}{0}
\pagebreak
\begin{figure} \begin{center} \epsfig{file=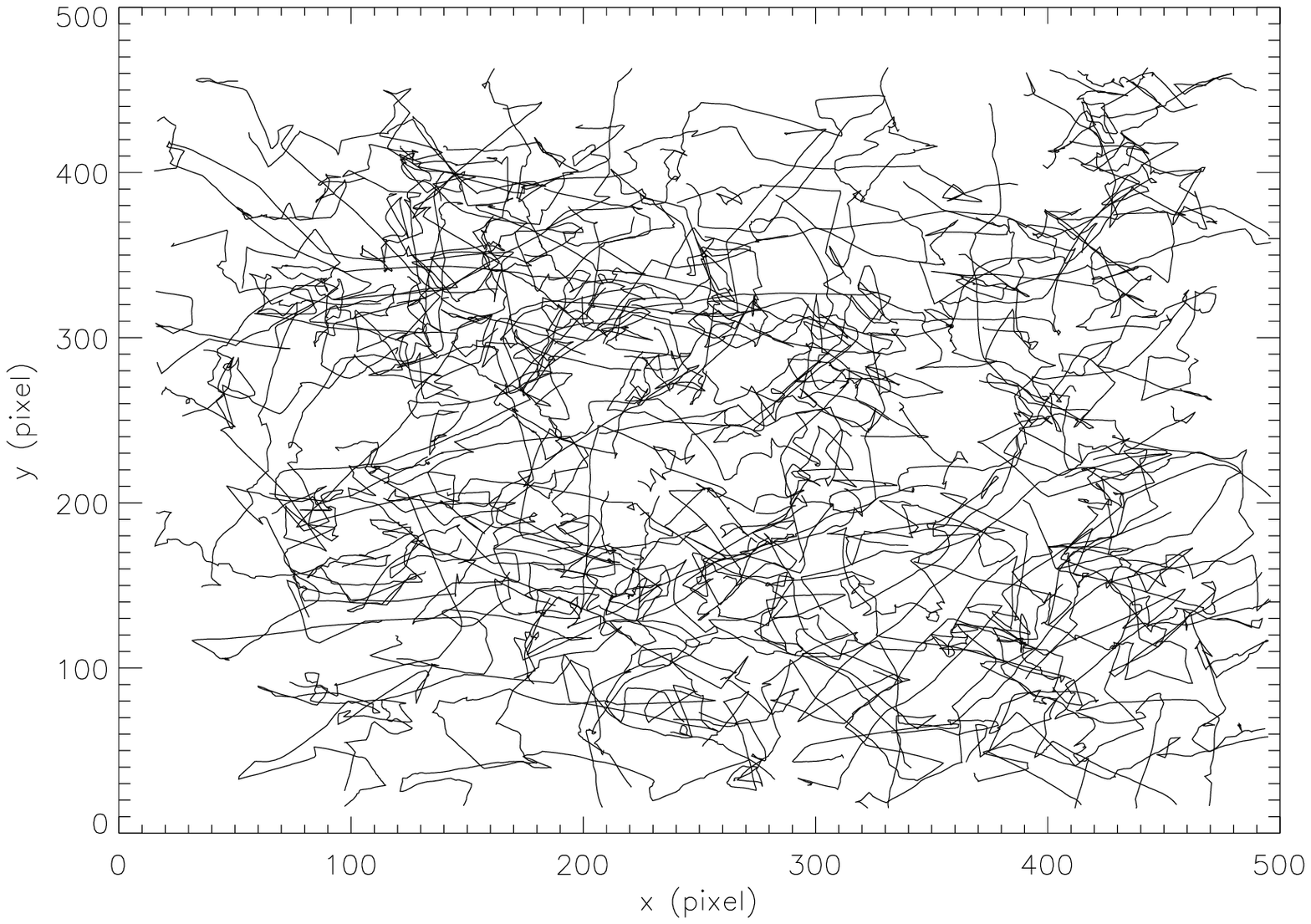,
width=\linewidth} \end{center} 
\caption{Measured trajectories of some  white particles 
moving among a sea of dark ones at a total coverage $c=0.28$. 
The container is vertically vibrated at 100 Hz and peak acceleration 
5 g. 
We acquire $546$~frames at $250$ frames/s. 
The particle diameter is about $11$~pixels).} 
\end{figure}

\begin{figure} \begin{center} 
\epsfig{file=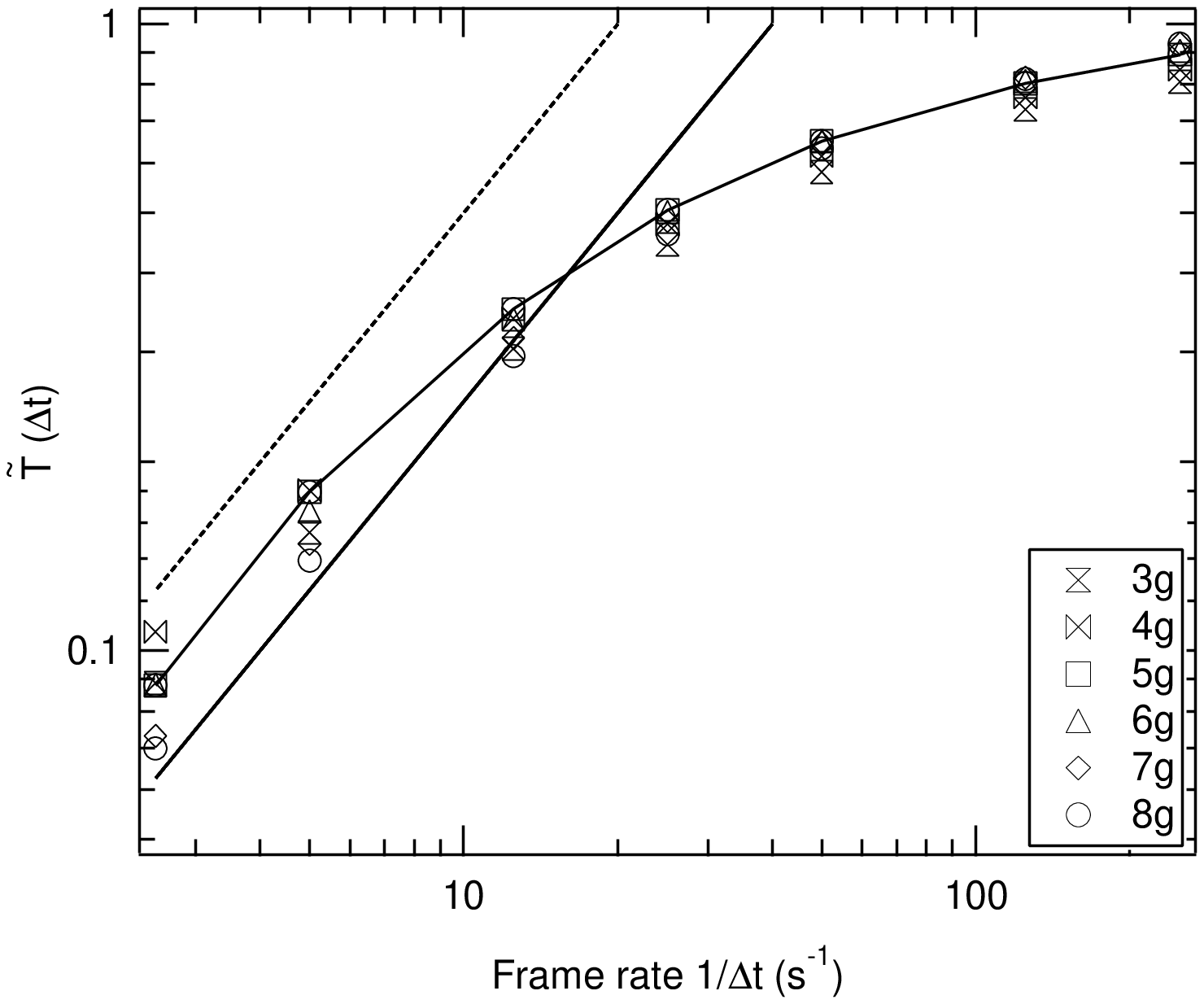, width=\linewidth} 
\end{center} 
\caption{Apparent non-dimensional granular temperature $\tilde{T}( \Delta t)$
corresponding to displacements along one axis
in the sample interval as a function of sampling rate $1/ \Delta t$.    
The approach to saturation for large sampling rates occurs when the
sampling rate far exceeds the collision rate. In the opposite limit
$\tilde{T}$ decreases strongly. The lines are explained in the text. }
\end{figure}

\begin{figure} \begin{center} 
\epsfig{file=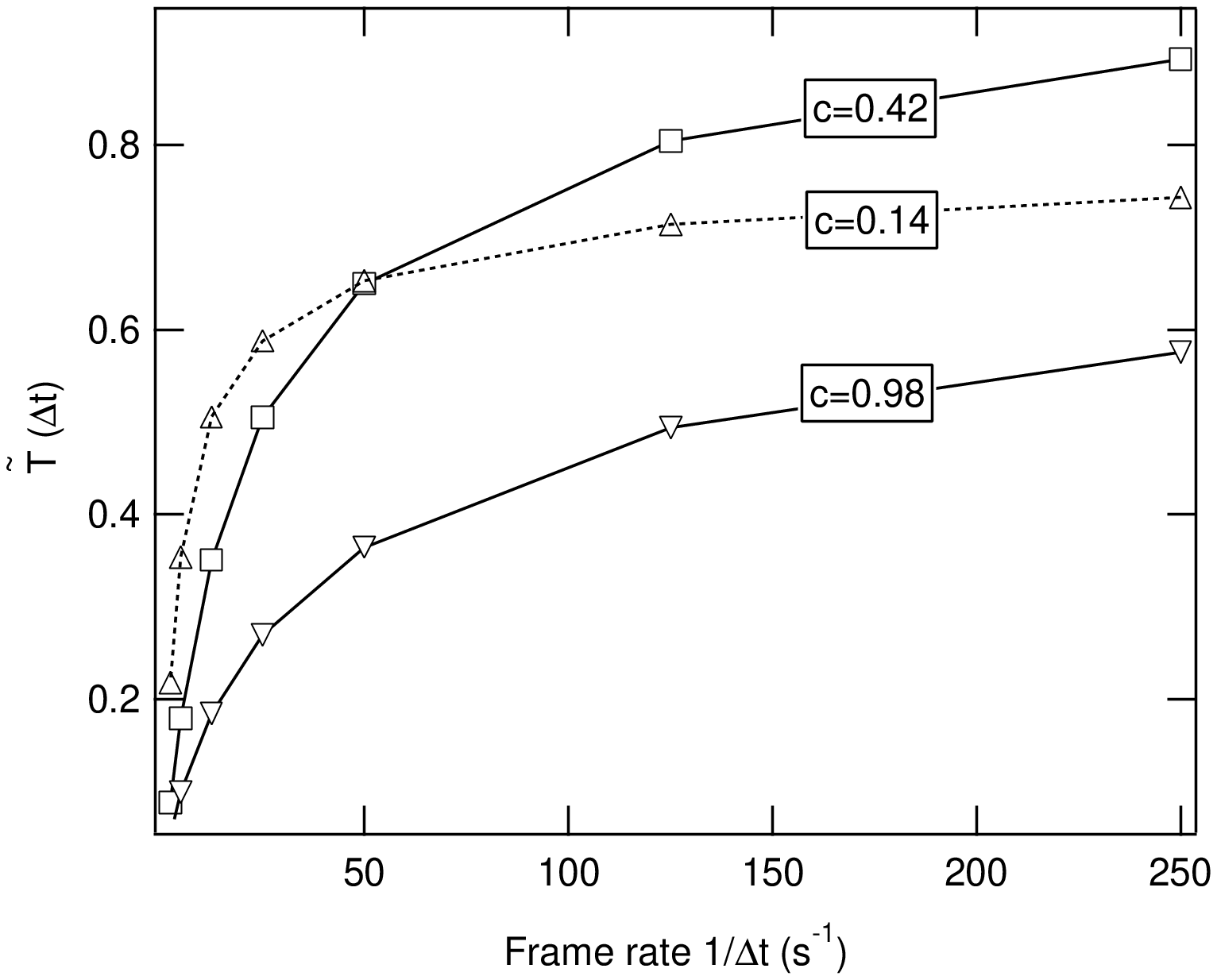, width=\linewidth} 
\end{center} \caption{Apparent non-dimensional granular temperature 
 $\tilde{T}(\Delta t)$ corresponding to displacements along one axis
in the sample interval $ \Delta t$
as a function of sampling rate $1/\Delta t$ for $a=5$ g. 
The limiting value for large sampling rates depends on coverage.}
\end{figure}

\begin{figure} \begin{center} 
\epsfig{file=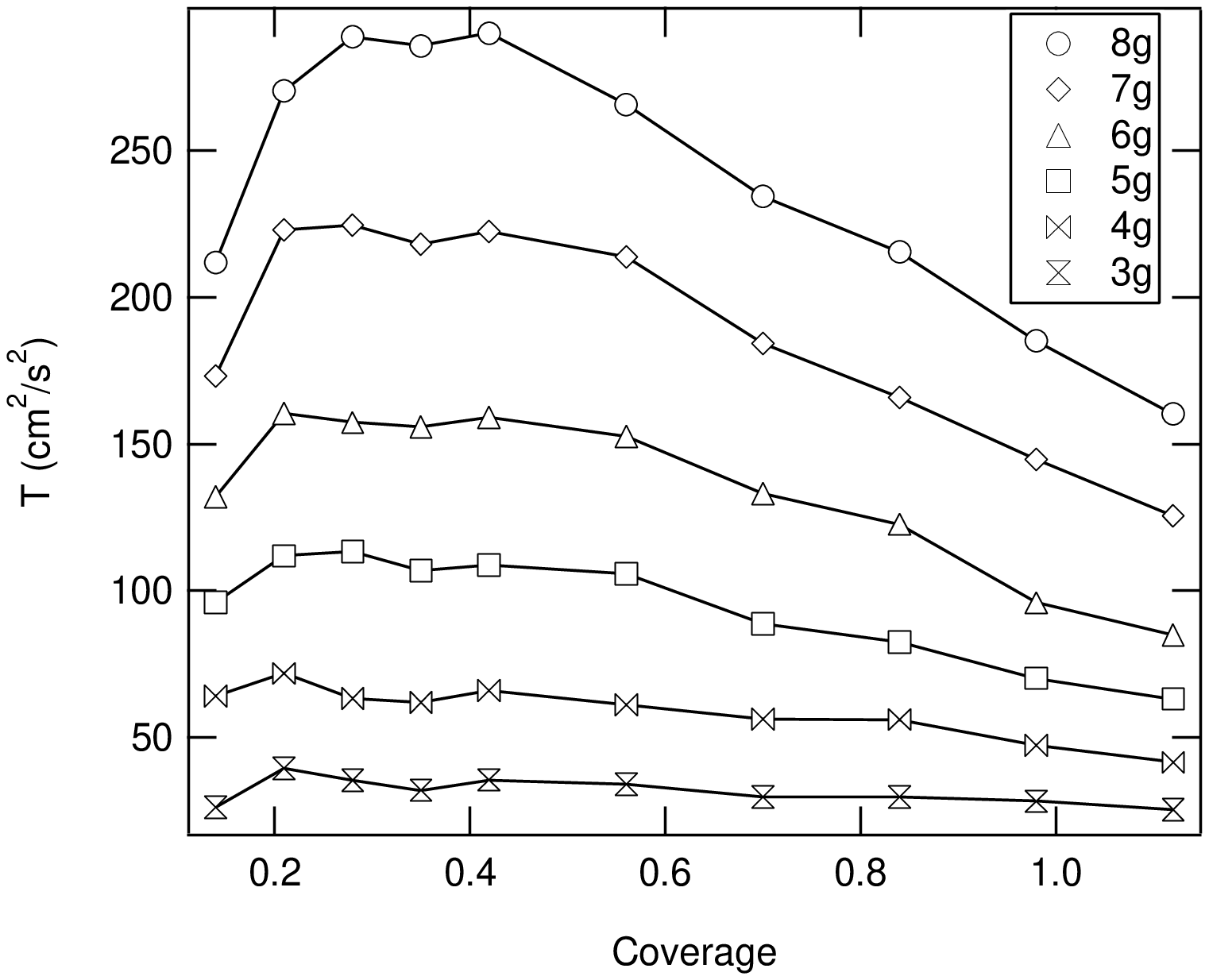, width=\linewidth} 
\end{center} \caption{Granular temperature $T$ determined 
from displacements along one axis as a function
of coverage for a range of peak plate accelerations.} 
\end{figure}

\begin{figure} \begin{center} 
\epsfig{file=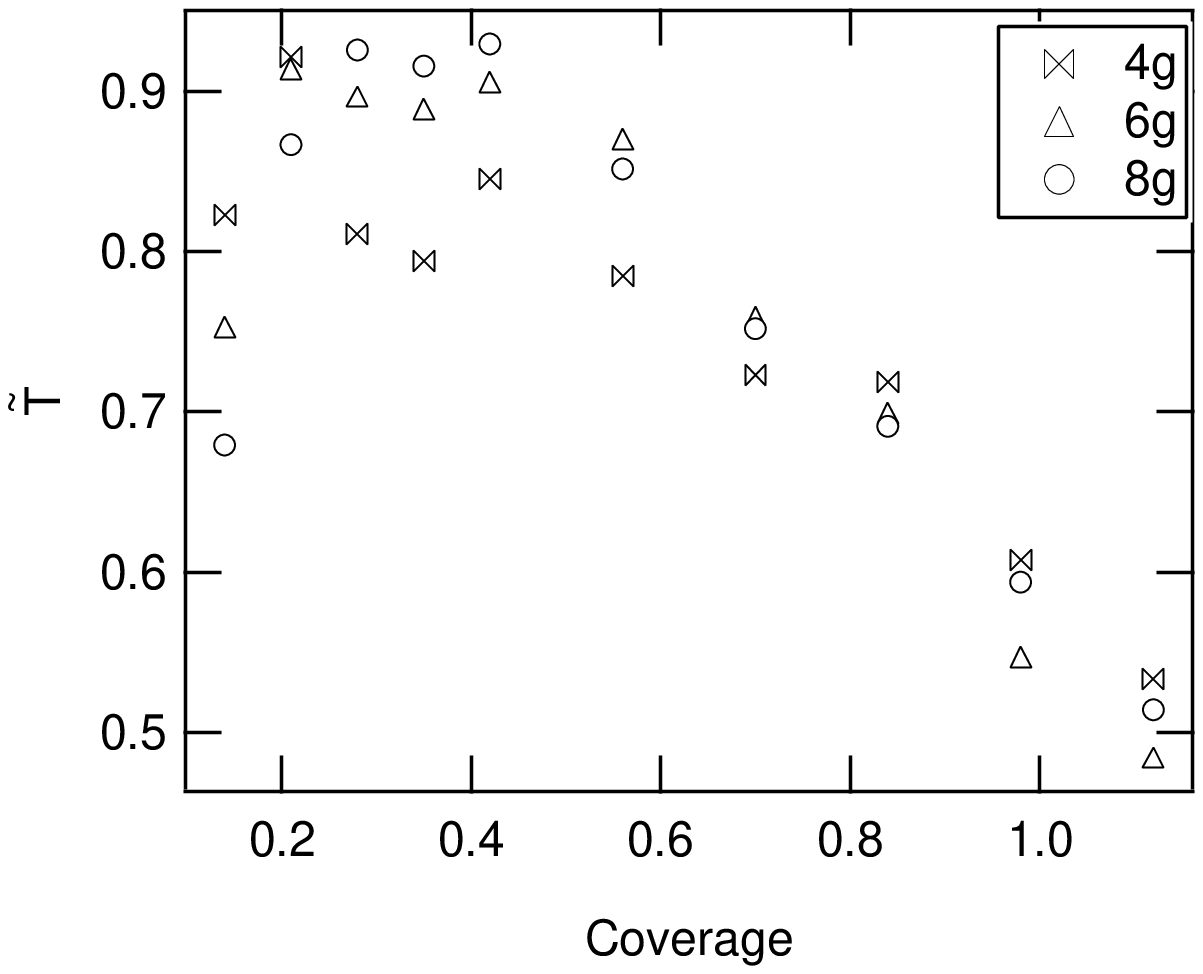, width=\linewidth} 
\end{center} \caption{Non-dimensional granular temperature $\tilde{T}$ 
vs. coverage. It is approximately independent of acceleration.}
\end{figure}

\begin{figure} \begin{center} 
\epsfig{file=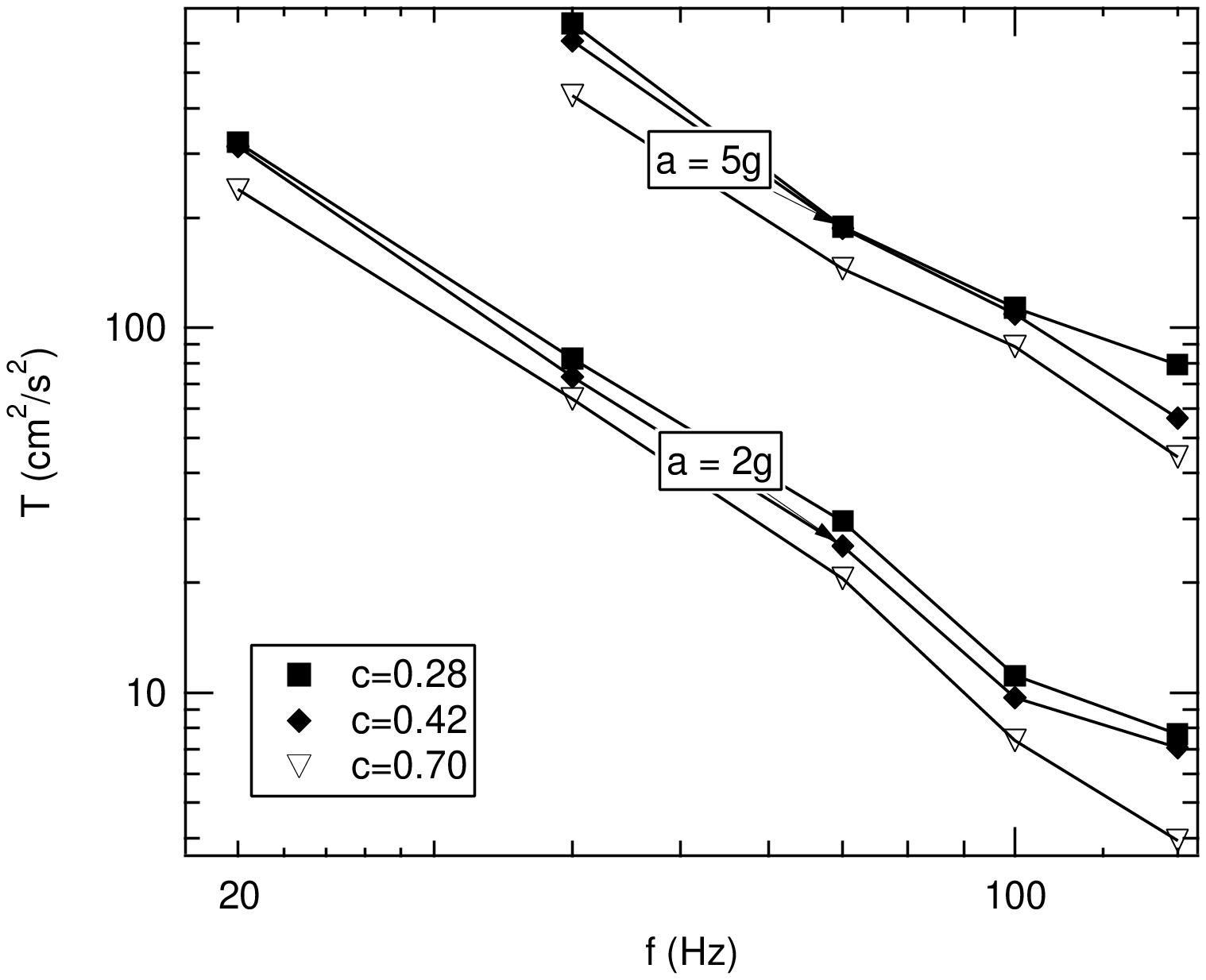, width=\linewidth} 
\end{center} \caption{Granular temperature $T$ vs.
vibration frequency $f$ at three coverages. It declines
approximately as $1/f^2$.}
\end{figure}

\begin{figure} \begin{center} 
\epsfig{file=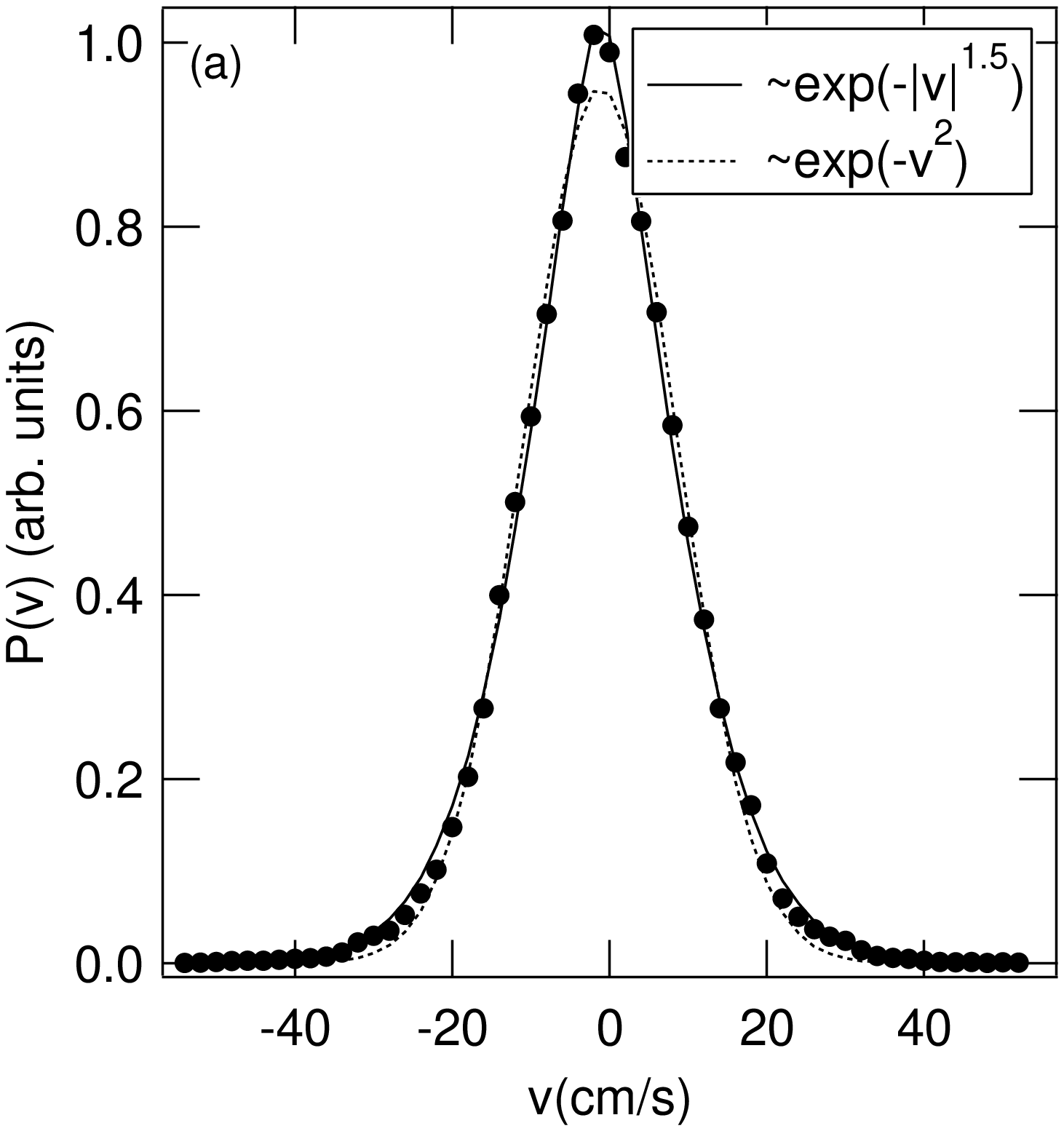, width=8 cm} \epsfig{file=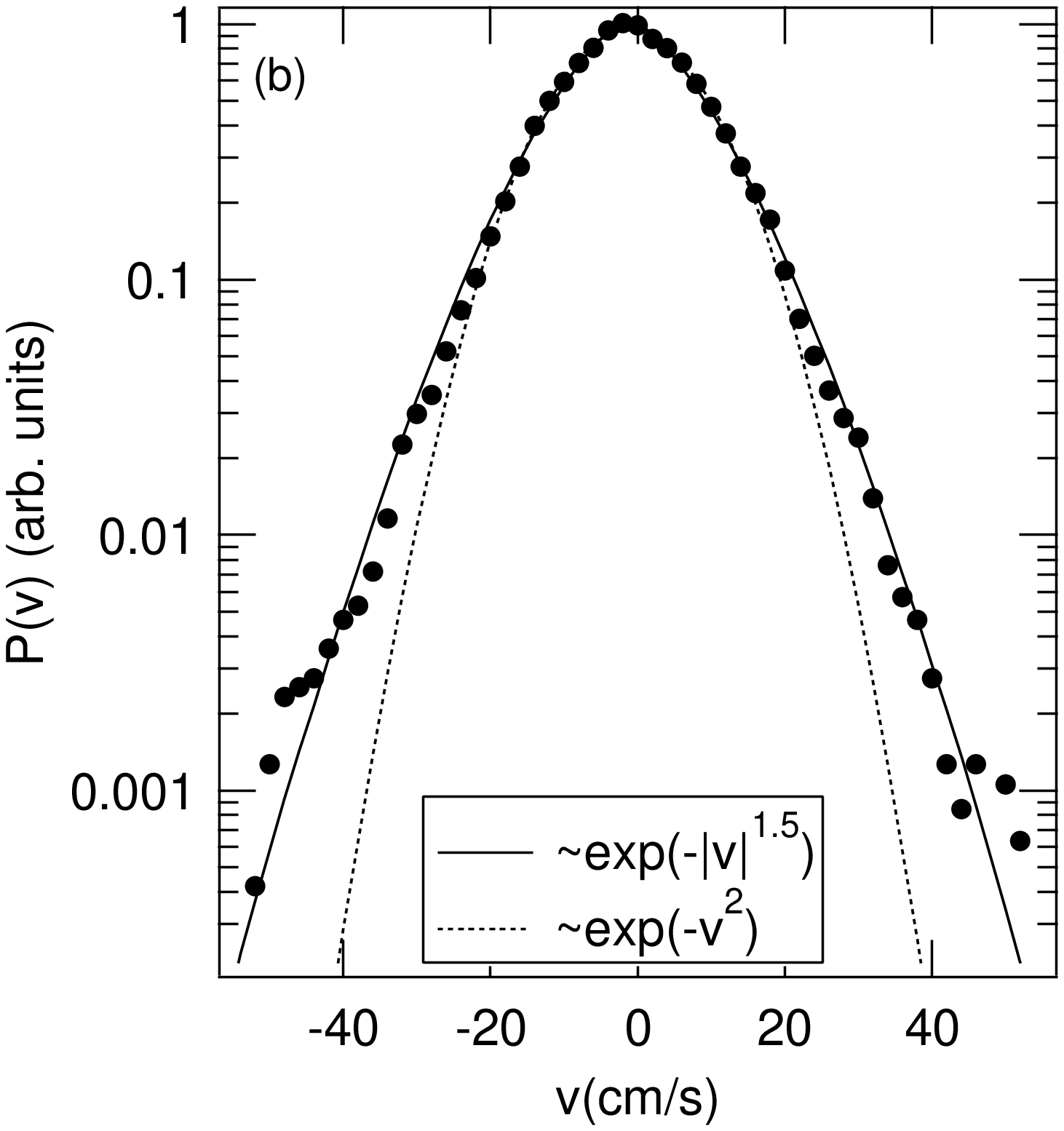, width=8 cm} 
 \epsfig{file=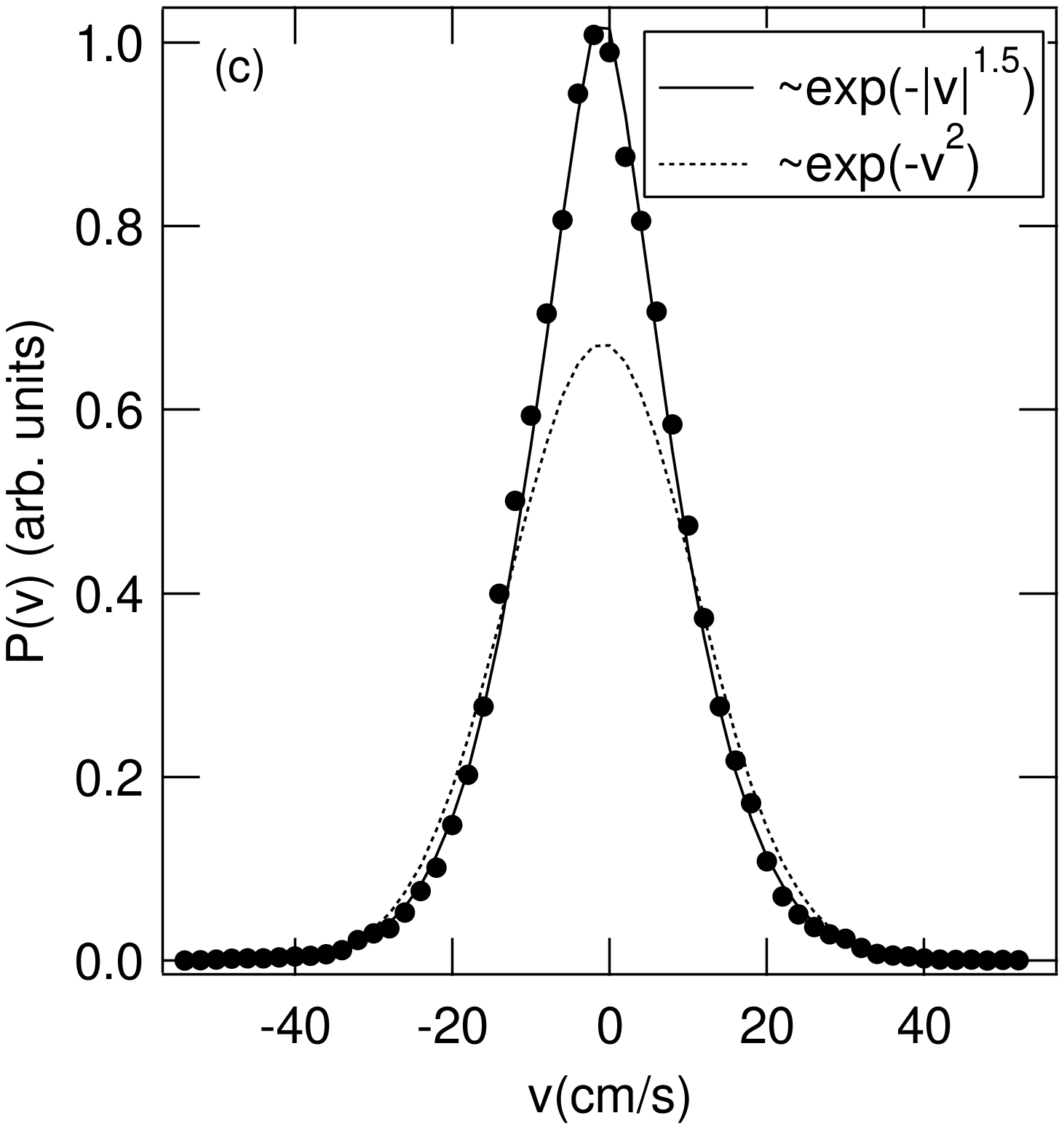, width=8 cm} \epsfig{file=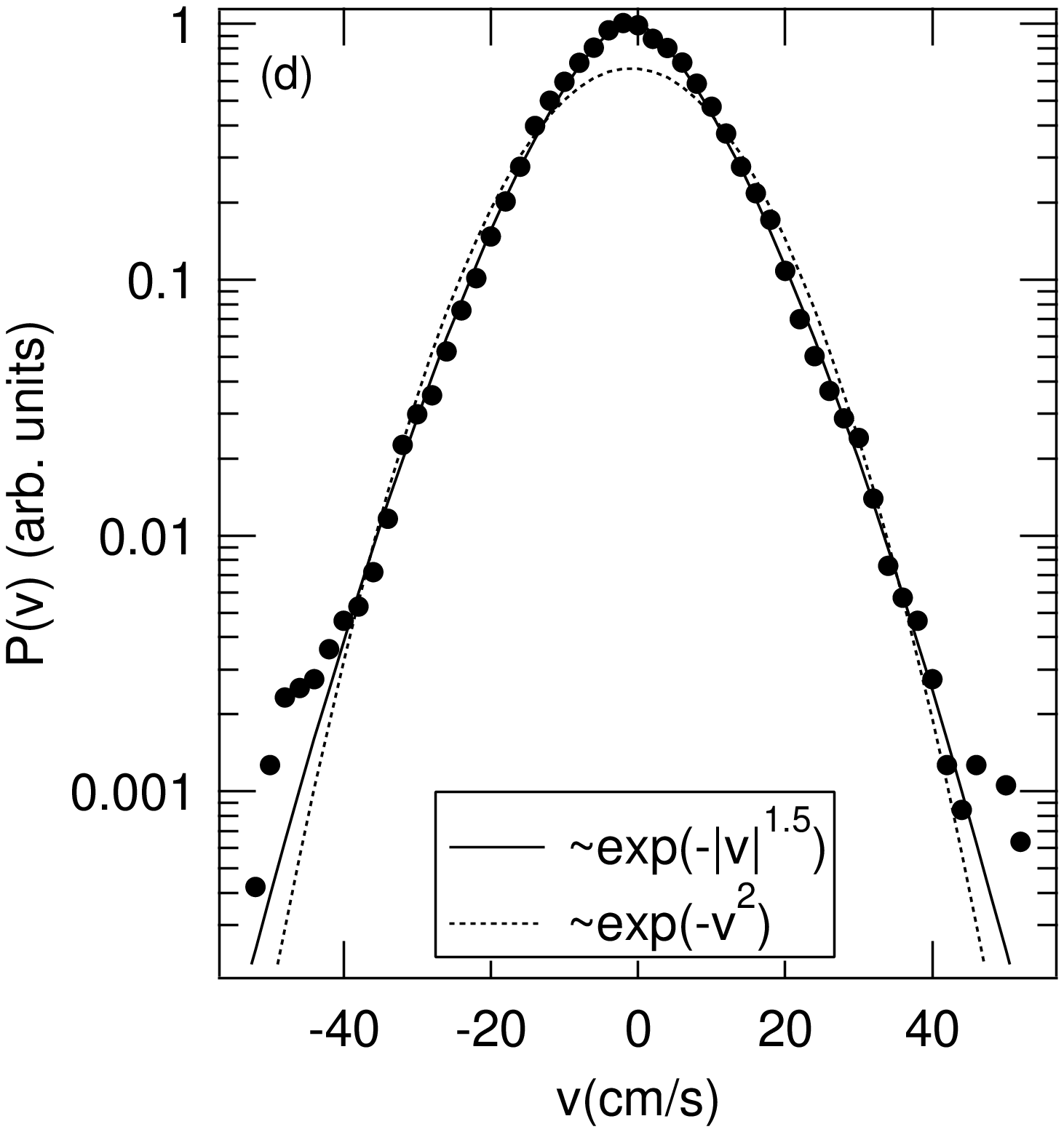, width=8 cm} 
\end{center}
\caption{Velocity distribution of the tracked particles 
plotted on linear (a,c) and logarithmic (b,d) scales (driving acceleration 
$a=5$ g, coverage $c = 0.42$, and frequency $f = 100$~Hz). The solid line is a 
fit to $F_2$ (Eq.~\ref{equnonGauss}).
The Gaussian fit 
(dashed line) 
underestimates the 
probablilities at high velocities. The data were weighted 
equally on linear scales in (a,b) and equally on logarithmic scales 
in (c,d), respectively.} 
\end{figure}

\begin{figure} \begin{center} 
\epsfig{file=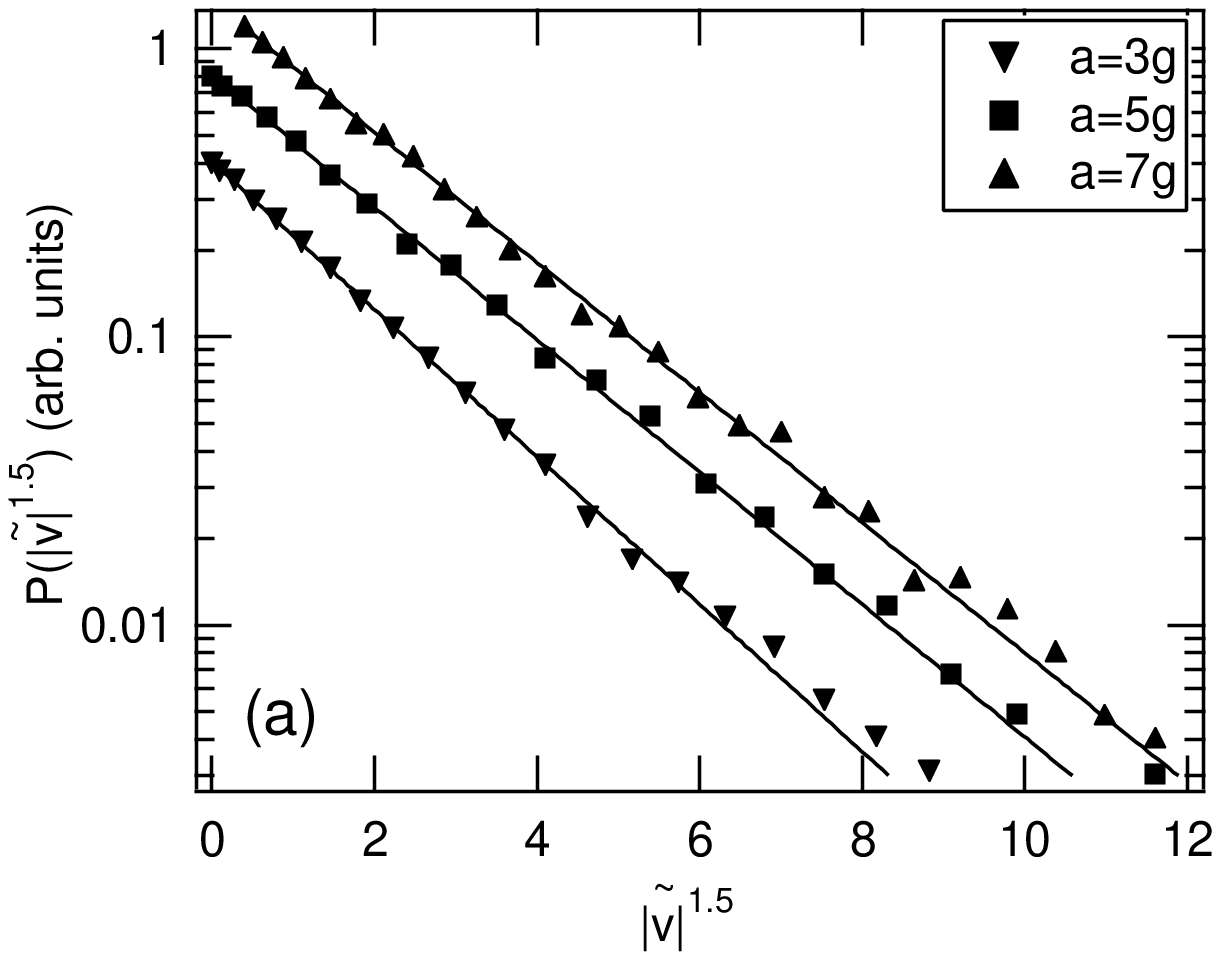, width=10cm} \epsfig{file=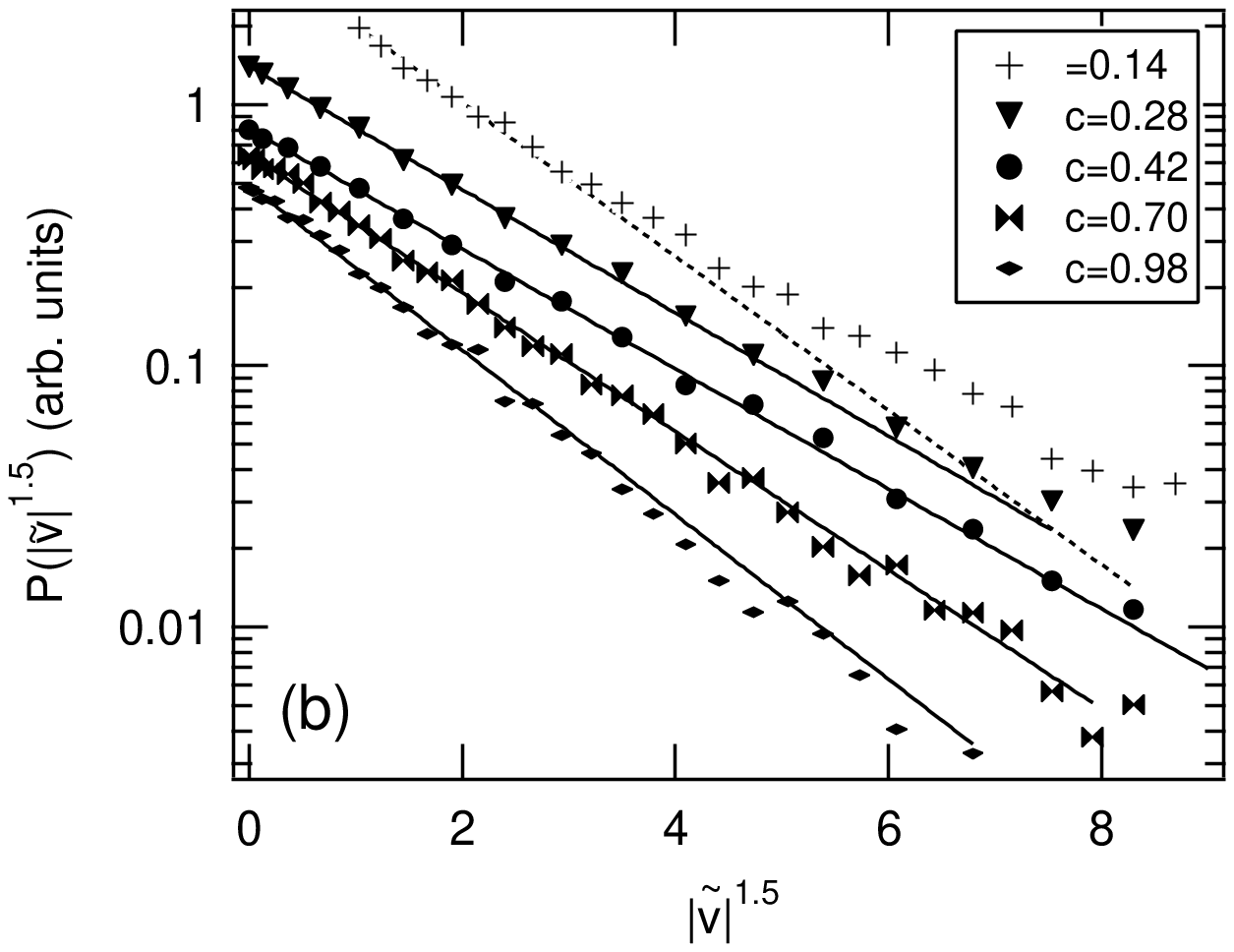, width=10cm} 
\end{center} \caption{Distributions of non-dimensional 
velocity $\tilde{v}$ obtained from
displacements along one direction
 vs. $\tilde{v}^{1.5}$ for (a) a range of accelerations and
(b) a range of coverages. Fits to Eq.~\ref{equnonGauss} (lines) are
also shown. Data are shifted vertically in some cases for clarity. Deviations 
from Eq.~\ref{equnonGauss} occur for $c\le 0.28$.}
\end{figure}

\begin{figure} \begin{center} 
\epsfig{file=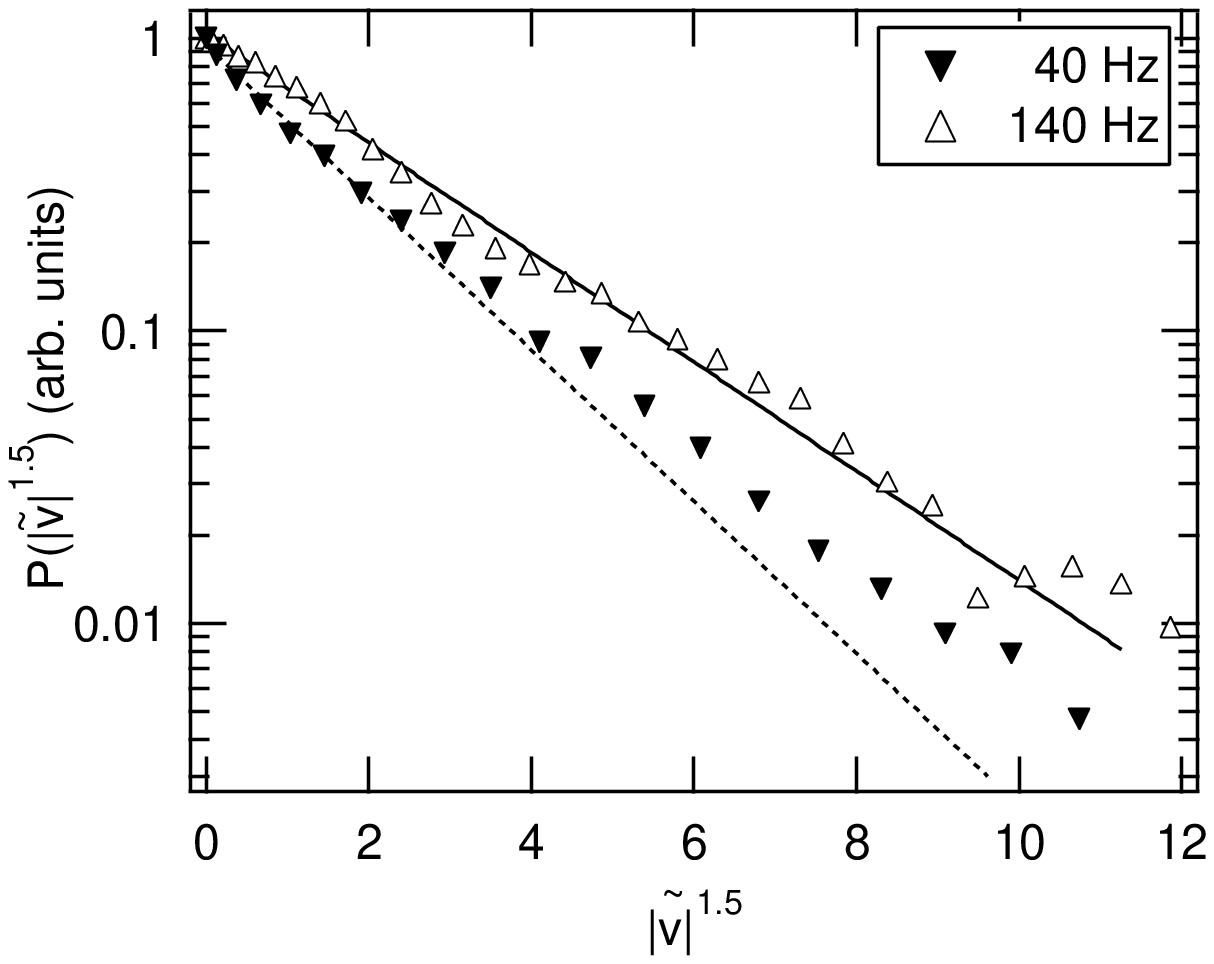, width=\linewidth} 
\end{center} \caption{Distributions of non-dimensional 
velocity $\tilde{v}$ obtained from
displacements along one direction vs.
 $\tilde{v}^{1.5}$ at different vibration frequencies,
and fits to Eq.~\ref{equnonGauss}.
($c=0.42$, $a=5g$)}
\end{figure}

\begin{figure} \begin{center} 
\epsfig{file=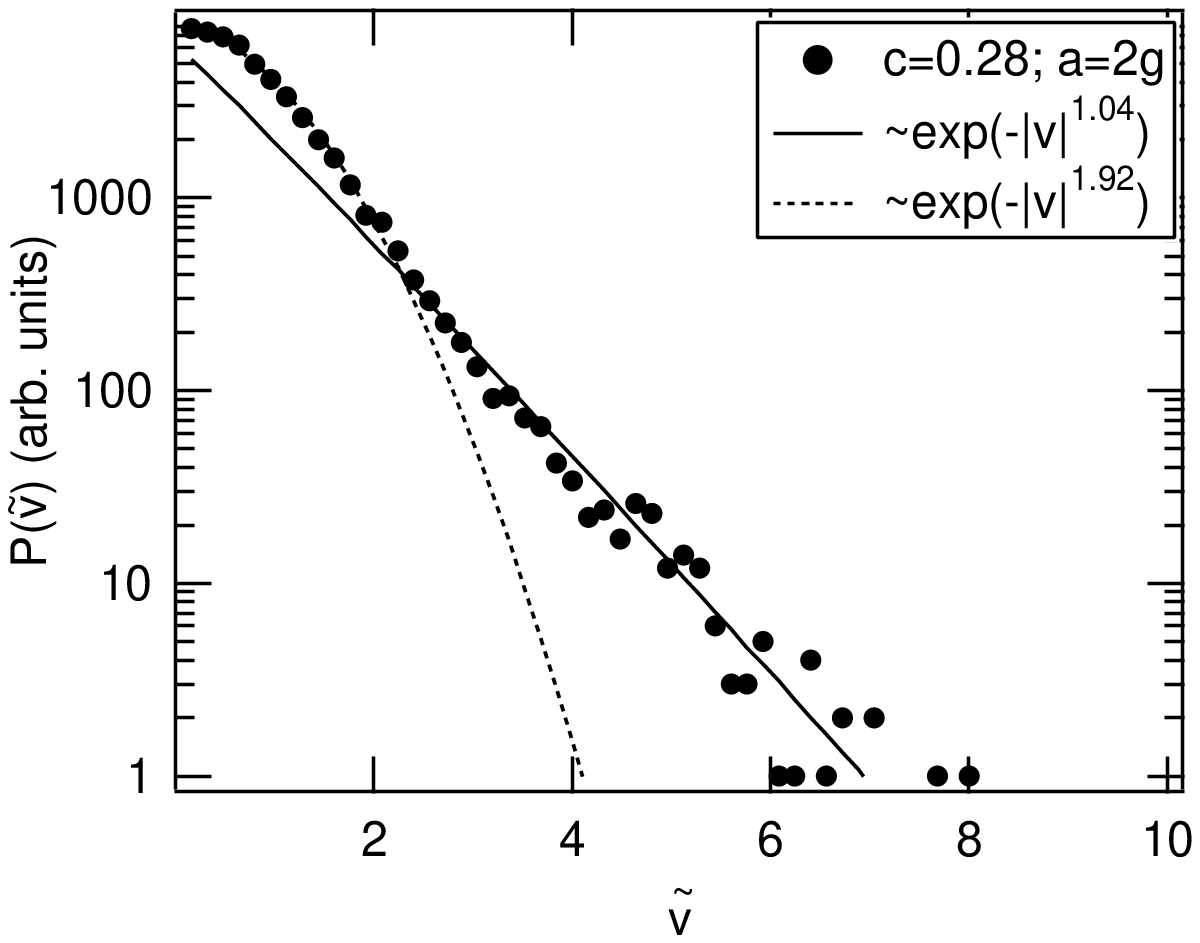, width=\linewidth} 
\end{center} \caption{Non-dimensional velocity distribution at 
low acceleration $a=2$~g, and fits to
$P(\tilde{v}) \sim exp(-|v|^\alpha)$ 
showing the crossover from a Gaussian to an exponential function.}
\end{figure}

\begin{figure} \begin{center} 
\epsfig{file=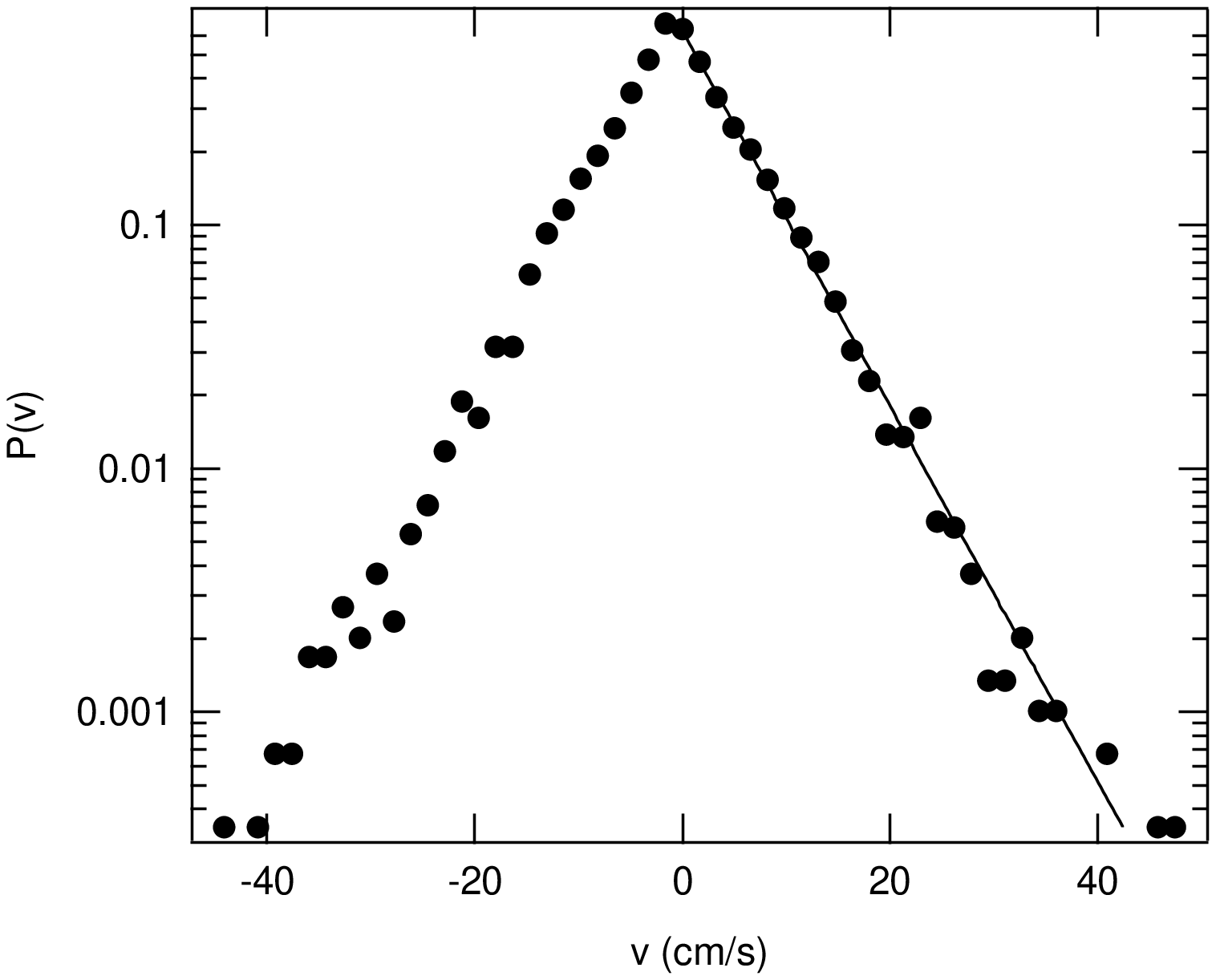, width=\linewidth} 
\end{center} \caption{Time averaged velocity distribution
during free cooling after excitation at $a=5$~g and $f=100$~Hz ($c=0.84$).
The high velocity tail is exponential 
(a straight line on this log-linear plot).}
\end{figure}

\begin{figure} \begin{center} 
\epsfig{file=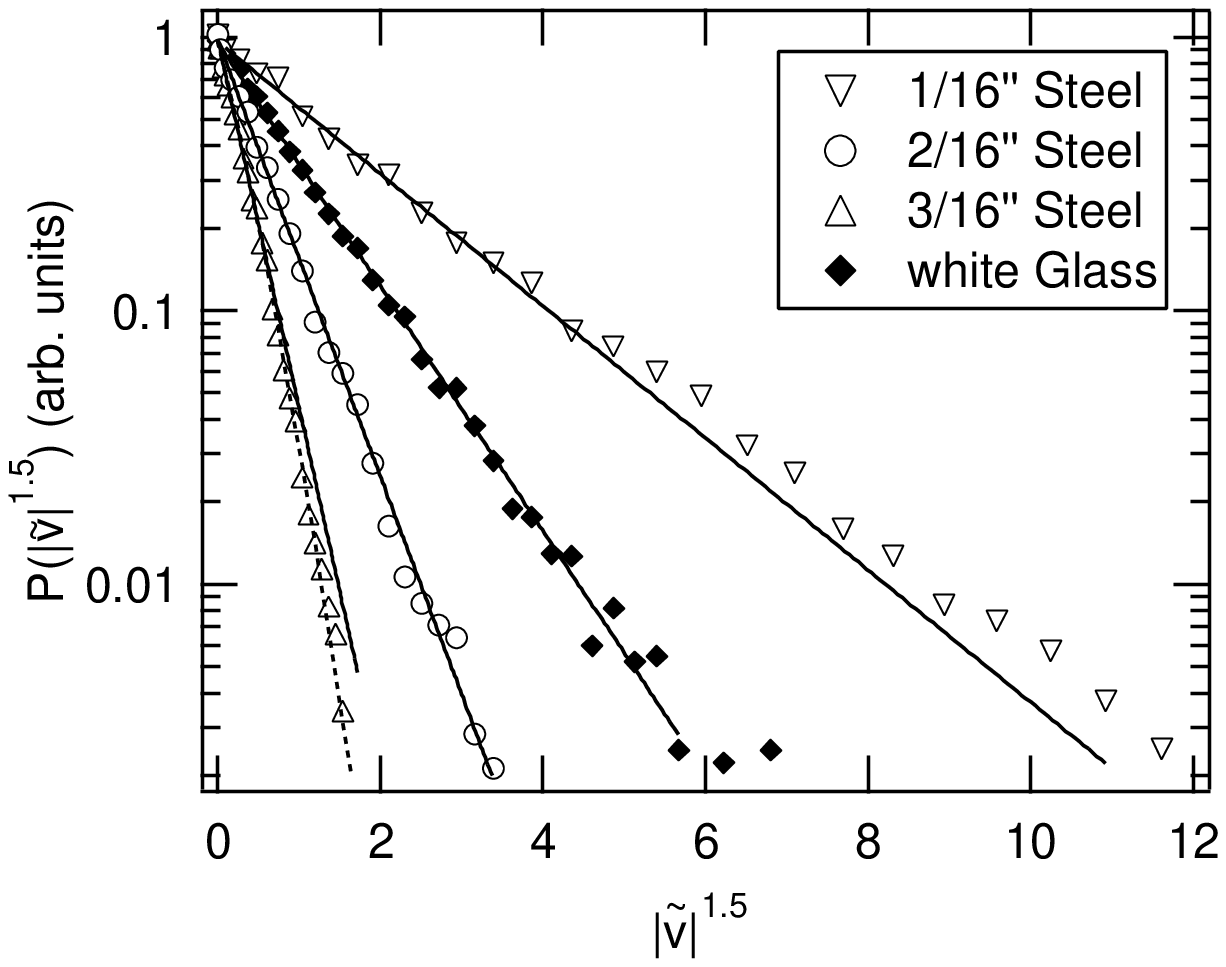, width=\linewidth} 
\end{center} \caption{Velocity distributions for steel beads moving among 
glass beads ($c=0.42$, $a=4$~g, $f=100$~Hz). Distributions are
fitted to Eq.~\ref{equnonGauss}
(solid lines) and also to a Gaussian (dashed line) for the largest beads.}
\end{figure}

\begin{figure} \begin{center} 
\epsfig{file=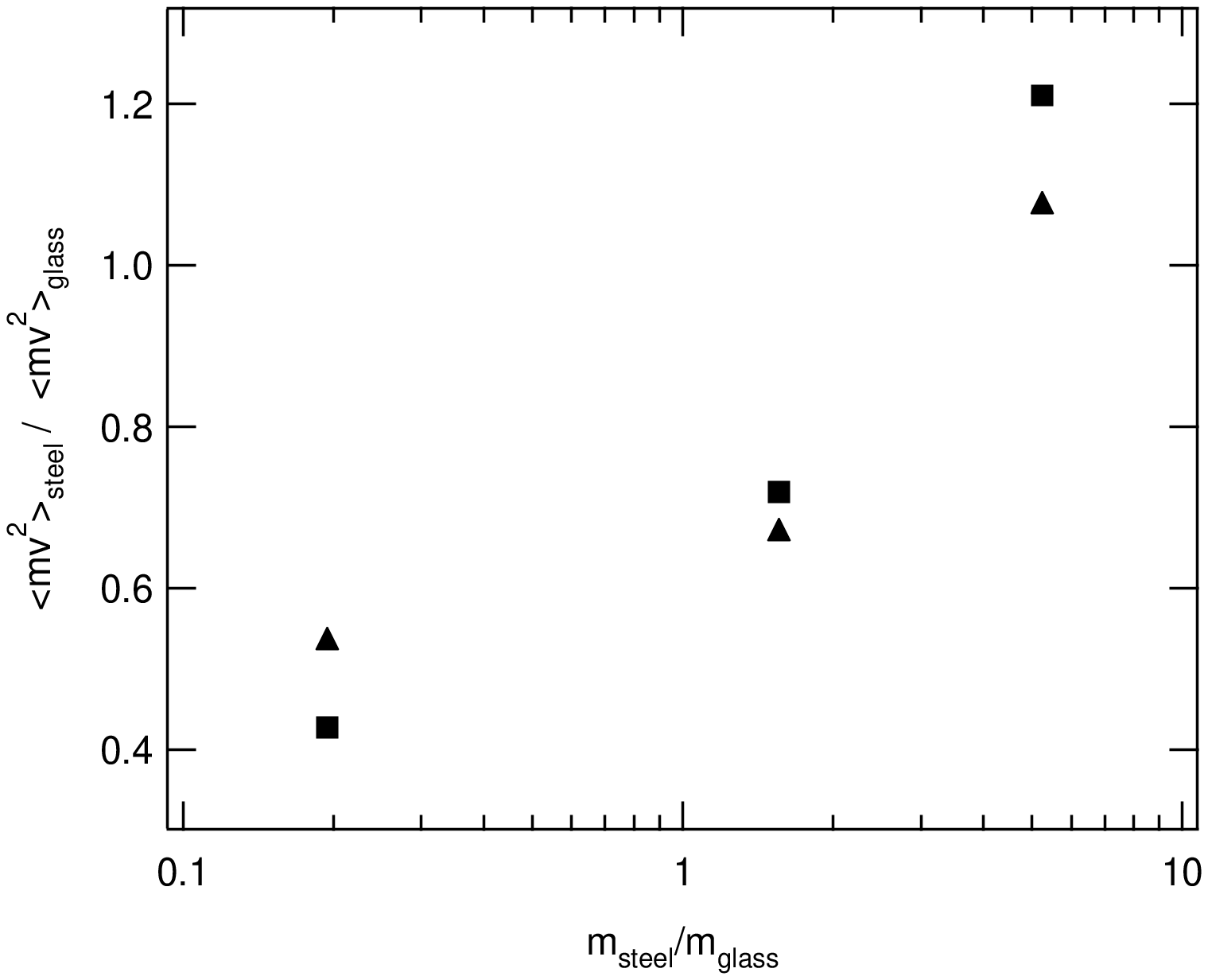, width=\linewidth} 
\end{center} \caption{Mean kinetic energy
of steel beads moving among glass beads (scaled by the energy of
the surrounding glass beads) as a function of particle mass.
($a=4$~g; $f=100$~Hz; squares, $c=0.42$; triangles $c=0.84$)}
\end{figure}

\end{document}